\documentclass[12pt]{article}

\usepackage[margin=1in]{geometry}
\usepackage{amsmath,amssymb}
\usepackage{graphicx}
\usepackage[style=numeric-comp,sorting=none,maxbibnames=3,minbibnames=1,giveninits=true,natbib=true,backend=biber]{biblatex}
\usepackage{booktabs}
\usepackage{float}
\usepackage{hyperref}
\usepackage{setspace}
\usepackage[title]{appendix}
\usepackage{tikz}
\usepackage[ruled,vlined]{algorithm2e}
\usetikzlibrary{arrows.meta,positioning,shapes,fit,calc}

\title{A model-based adaptive sampling framework for infectious disease surveillance}

\author{
Supun Manathunga$^{1}$, Nathaniel Osgood$^{2}$, Sheila F O’Brien$^{3,4}$,\\
David L Buckeridge$^{1,5}$, W. Alton Russell$^{1,5}$ \\
\\
\small{$^1$Division of Experimental Medicine, McGill University, Montreal, Canada} \\
\small{$^2$Department of Computer Science, University of Saskatchewan, Saskatoon, Saskatchewan, Canada} \\
\small{$^3$Canadian Blood Services, Ottawa, Ontario, Canada} \\
\small{$^4$School of Epidemiology and Public Health, University of Ottawa, Ottawa, Ontario, Canada} \\
\small{$^5$School of Population and Global Health, McGill University, Montreal, Canada} \\
}

\date{}

\begin{document}

\maketitle

\begin{abstract}
\noindent\textbf{Background:}
Longitudinal surveillance can monitor infectious disease dynamics over time, but each surveillance round requires a sampling design to be chosen before the next observations are available. Conventional fixed sample-size rules treat each round as an isolated cross-sectional survey and do not account for the evolving uncertainty in the epidemic process. In an epidemic, this uncertainty can change substantially over time as transmission, vaccination, public health interventions, pathogen variants, and behavior change.

\noindent\textbf{Objective:}
We present a model-based adaptive sampling framework for determining the next-round sampling design for longitudinal infectious disease surveillance. Applying the framework retrospectively to SARS-CoV-2 serosurveillance data for four regions of Canada, we assess its ability to select the minimum weekly sample size needed to develop sufficiently precise seroprevalence estimates.

\noindent\textbf{Methods:}
The framework has three components. First, a state-space model represents the epidemic process and maps latent epidemic states to surveillance estimand(s). Second, a statistical filter sequentially calibrates the model as surveillance observations arrive, propagating uncertainty in latent states, time-varying parameters, stochastic epidemic dynamics, and sampling noise. Third, a Bayesian assurance criterion converts the one-step-ahead predictive distribution into the minimum-cost sampling design likely to achieve a target posterior credible-interval width. We evaluated the framework in two simulations and in a retrospective case study of SARS-CoV-2 serosurveillance among Canadian blood donors. The case study used a stratified stochastic transmission model and replayed the surveillance series, subsampling each week's tested specimens to the design the adaptive policy selected.

\noindent\textbf{Results:}
In simulations, the adaptive policy increased sampling effort during an anticipated high-variance period, and the share of intervals meeting the width target (0.970) agreed with validated assurance (0.974).
After an unanticipated change in transmission rate, validated assurance stayed at 0.960 while the share of intervals meeting the width target fell to 0.894 (95\% Monte Carlo interval 0.869--0.919) in the transition window. The policy raised sampling effort while the filter relearned the transmission rate and reduced it once the new rate was learned.
In the SARS-CoV-2 case study, recommended sample sizes tracked the one-step predictive uncertainty and were largest during the Omicron wave, and the 90\% filtering interval had width at most 0.10 in 94\% of region-rounds with specimens.

\noindent\textbf{Conclusions:}
Model-based adaptive sampling can adjust surveillance effort responsive to the evolving uncertainty in an epidemic. This modular framework can be adapted to other pathogens, data streams, and public health surveillance objectives.

\end{abstract}

\newpage

\section{Introduction}

Serial  or continuous surveillance plays a critical role in understanding infectious disease dynamics. Surveillance modalities include direct pathogen testing, case or syndromic surveillance, vector monitoring, and serosurveillance (testing for markers of immunity). These modalities differ in what they measure, but each must repeatedly decide how much sampling effort the next round requires.

The SARS-CoV-2 pandemic drove unprecedented investment in repeated serological surveys, but the infrastructure and funding for large-scale testing are difficult to sustain. Future surveillance systems therefore need designs that preserve inferential precision while using sampling effort efficiently.

Assume collection of samples for surveillance happens in sequential rounds. At round $t$, one must choose the sampling design for round $t+1$ before the round-$t+1$ data are observed. The design sets how many samples to collect and usually how that total is distributed across strata such as age, sex, or region. The target estimand, such as infection-induced seroprevalence, is itself changing because population immunity is affected continually by infection, vaccination, waning, and pathogen evolution \citep{metcalf_serological_2016,muller_learning_2023}. The uncertainty around the estimand may also vary. In some periods, previous observations and mechanistic constraints tightly determine the near-future trajectory while in others the epidemic state is poorly constrained \citep{flaxman_estimating_2020,brauner_inferring_2021,chang_mobility_2021,volz_assessing_2021,carabelli_variant_2023}. A useful adaptive design should increase sampling when the trajectory is genuinely uncertain and reduce sampling when it is already well constrained.

Conventional sample-size rules are poorly matched to this problem. Fixed margin-of-error calculations depend on an assumed prevalence to calculate sample size for a single cross-sectional survey \citep{fosgate_practical_2009}. A systematic review of early COVID-19 prevalence surveys found highly heterogeneous sample sizes, flagged sample size as a major risk-of-bias domain, and noted that rounds often required different sample sizes as prevalence changed, with some studies under-sized in some rounds \citep{franceschi_population-based_2021}.

Bayesian sample-size designs select a sample size by averaging over the predictive distribution of the unseen data and asking whether the resulting posterior will be precise enough. A design prior fixed at the planning stage drives the predictive distribution, so these methods do not extend directly to settings where the estimand is a `moving target' resulting from a dynamic process, such as a disease epidemic. Statistical filtering, the sequential assimilation of observations into a state-space model, calibrates state-space models online (i.e., as data become available) and propagates uncertainty forward as new observations arrive \citep{daza-torres_bayesian_2022,gordon_novel_1993,doucet_introduction_2001}. These filters have been applied to infectious disease models for nowcasting, forecasting, and parameter updating as data become available \citep{dukic_tracking_2012,yang_inference_2014,he_plug-and-play_2010,funk_real-time_2018,dureau_capturing_2013}. But statistical filtering and Bayesian sample-size methods have rarely been combined, and the potential use of filtering predictive distributions as design priors for online sample-size determination has not been previously explored.

We propose a model-based adaptive sampling framework that links Bayesian sample-size determination and statistical filtering, treating selection of the next-round sampling design as a sequential Bayesian design problem. A state-space model represents the evolving epidemic and relates surveillance data to the system state to define the surveillance estimand. A statistical filter performs online model calibration, providing a one-step-ahead predictive distribution for the next round. A Bayesian assurance criterion uses the predictive distribution to derive the minimum-cost sampling design that meets a posterior precision target with a sufficient predictive probability.

The remainder of the paper first reviews Bayesian sample-size determination and particle filtering. It then presents the adaptive sampling framework in general form, evaluates its operating behavior in two simulations, and applies it to Canadian SARS-CoV-2 serosurveillance. The final sections report the retrospective evaluation and discuss the implications and limitations of the framework.

\section{Background and related work}
\label{sec:background}

\subsection{The decision-theoretic basis of sampling design}
\label{sec:bg-decision}

The idea that an experiment can be valued by the information it is expected to
provide draws on Shannon's entropy. For a discrete probability distribution
$p=(p_1,\ldots,p_m)$, \citet{shannon1948} defined entropy as
\[
H(p)=-\sum_{i=1}^{m}p_i\log p_i,
\]
which measures uncertainty about the outcome. Let $\theta$ denote an unknown
parameter and $Y$ the random outcome of an experiment. \citet{lindley1956}
defined the expected information provided by the experiment as
\[
I(\theta;Y)
=
\mathbb{E}_{Y}\mathbb{E}_{\theta\mid Y}
\left[
\log\frac{p(\theta\mid Y)}{p(\theta)}
\right].
\]
This quantity is the expected reduction in entropy from the prior to the
posterior, now known as the mutual information between $\theta$ and $Y$.
Originally, Lindley proposed using $I(\theta;Y)$ as a criterion when evaluating experimental designs, not as a measure of utility within a formal decision problem. He later applied the criterion to compare binomial
sampling schemes and to examine the relationship between information and inferential accuracy.

Independently, \citet{savage1954} established an axiomatic foundation for
ranking decisions by expected utility, later used by \citet{raiffa1961} to develop an early
and influential fully Bayesian treatment of experimental design as a sequential
decision problem. Their binomial example concerned a ``terminal act'' with economic consequences, accepting or rejecting a lot of manufactured items after inspecting a sample of them for defectives, rather than making inference about an uncertain parameter. In
1979, \citet{bernardo1979} connected Raiffa and Schlaifer's Bayesian decision
theory with the information measures of Shannon and Lindley, and showed
that, under mild smoothness conditions, a utility for reporting a probability
distribution that is both \emph{proper} and \emph{local} must have a logarithmic
form, placing Lindley's expected
information gain within Bayesian decision theory as a measure of expected utility.
Work of this nature continues in Bayesian experimental design and the
expected-information-gain literature \citep{chaloner1995,rainforth2024}.

Decision-theoretic methods are coherent and can accommodate many utility
functions, but their implementation can be difficult. When sample size is chosen
by maximizing expected utility with a sampling cost, the analyst must express the
value of the resulting inference and the cost of sampling in the same units (e.g., in dollars)
\citep{lindley1997}. Explicit utility functions are difficult to define in infectious disease
surveillance because the ways in which information will improve health are not
always known in advance. These practical difficulties explain the appeal of
methods that replace an explicit utility and cost trade-off with directly
interpretable constraints on the precision of the final estimate.

\subsection{Precision-based Bayesian sample-size determination}
\label{sec:bg-precision}

In 1997, \citet{lindley1997} compared maximization of expected utility with
interval-based criteria that had recently been proposed for binomial estimation
\citep{joseph1995,adcock1995}. Lindley regarded these criteria as not fully
Bayesian because they imposed constraints on posterior coverage and/or interval
length rather than balancing inferential utility against sampling cost. He showed
that an indicator for interval membership could incorporate coverage
criteria into a decision-theoretic formulation, but such formulations were not equivalent to maximization of expected utility. He also argued
that the average length criterion could violate coherence. Yet constraint-based criteria have practical appeal, since constraints such as interval coverage and
length are easier to elicit from domain experts than utility functions.

\citet{joseph1995} introduced three criteria based on highest posterior density
intervals. Before data are collected, the posterior probability mass of
an interval and its length depend on
the dataset yet to be observed. The \emph{average coverage criterion} (ACC)
fixes the length and requires the expected probability mass over the prior
predictive distribution to reach a target. The \emph{average length criterion}
(ALC) fixes the probability mass and constrains the expected length. The \emph{worst outcome
criterion} (WOC) requires both targets to hold for every possible dataset, and
the \emph{modified worst outcome criterion} (MWOC) of \citet{joseph1997}
relaxes this to a high-probability subset of the sample space.
\citet{mlan2006} added median-based versions of these criteria, and
\citet{mlan2008} compared them for binomial sample-size determination.

\citet{ohagan2001,ohagan2005} introduced the concept of assurance separately in clinical trial design as an alternative to power. Power conditions on an assumed true parameter
value, whereas assurance averages the probability of study success over a design
prior. \citet{wilson2022} applied assurance to estimating the sensitivity and specificity of a diagnostic test by defining success as
a posterior interval no wider than a target $w^*$. Let $W(Y,n)$ be the width of
the posterior credible interval from a future dataset $Y$ of size $n$. The
assurance at sample size $n$ is
\[
A(n)=\Pr\!\left\{W(Y,n)\le w^*\right\},
\]
with the probability taken over the prior predictive distribution of $Y$, and
the sample size is chosen so that $A(n)\ge q$ for a required assurance level
$q$. This constrains a high quantile of the predictive width
distribution rather than its mean and Section~\ref{sec:framework} adapts this
criterion to repeated surveillance.

\subsection{From fixed designs to adaptive surveillance}
\label{sec:bg-adaptive}

Sequential and adaptive Bayesian trial designs use accumulating information and revise interim decisions as the data arrive, but they target a static estimand, typically a fixed treatment effect \citep{jennison_group_1999,muller_simulation_2007,berry_bayesian_2010}.
Surveillance estimands are instead time-varying functions of a latent process.
Let $Y_s$ denote the surveillance observations in round $s$ and $D_s$ the
sampling design used in that round, with realized values $y_s$ and $d_s$.
The surveillance history after round $t$ is the random vector
$H_t=(Y_{1:t},D_{1:t})$, with realized value $h_t=(y_{1:t},d_{1:t})$.
Conditioning on $h_t$ means conditioning on the event $H_t=h_t$. At the end of
round $t$, the design $d_{t+1}$ for the next round must be chosen from $h_t$
alone, before $Y_{t+1}$ is observed.

The interval-based criteria of Section~\ref{sec:bg-precision} extend to this
setting by requiring that the round-$(t+1)$ posterior interval meet a target
width with high probability. The probability is taken over the predictive
distribution of $Y_{t+1}$ given $h_t$, which changes from round to round. Let
$x_t$ denote the latent state of the process and $\psi_t=\varphi(x_t)$ the
surveillance estimand. The criterion then requires several related predictive
distributions. A statistical
filter updates the filtering distribution $p(x_t\mid h_t)$ and propagates
it through the process model to obtain the next-round state predictive
distribution $p(x_{t+1}\mid h_t)$. Mapping the predicted state through
$\varphi$ gives the predictive distribution of $\psi_{t+1}$. Combining the state predictive distribution with the
observation model under a candidate design gives the design-dependent
predictive distribution of the next observation, which is the future-data
distribution required by the assurance calculation. Statistical filtering therefore provides the evolving filtering and state
predictive distributions.

\subsection{The filtering problem and Bayesian recursion}
\label{sec:bg-filtering}

In a state-space model,
an unobserved state $x_t$ evolves as a Markov process with transition density
$p(x_{t+1}\mid x_t)$, and an observation $y_t$ is generated from the current
state through an observation density $p(y_t\mid x_t,d_t)$. In serosurveillance, $x_t$ contains the latent epidemic state and its
time-varying parameters, while $y_t$ may contain the round-$t$ serological counts.

\citet{kalman_new_1960} developed a recursive optimal linear filter using a
state transition representation of the process, which extended
the earlier spectral formulation of \citet{wiener_extrapolation_1949}, and provides the exact filtering distribution under
linear and Gaussian assumptions. \citet{ho_bayesian_1964} subsequently formulated
multistage state estimation in explicitly Bayesian terms and showed how the full
posterior density could be propagated and updated recursively for nonlinear and
non-Gaussian models.

At the end of round $t$, the filtering distribution is
$p(x_t\mid h_t)$ and propagating this distribution through the state
transition gives the prediction step
\begin{equation}
p(x_{t+1}\mid h_t)
=
\int p(x_{t+1}\mid x_t)\,p(x_t\mid h_t)\,dx_t.
\label{eq:bg-predict}
\end{equation}
Suppose that the next design is conditionally independent of the next state given
the available surveillance history,
\begin{equation}
D_{t+1}\mathrel{\perp\!\!\!\perp}x_{t+1}\mid H_t.
\label{eq:bg-noninformative-design}
\end{equation}
This condition holds when the design is fixed in advance, chosen as a function of
$ H_t$, or is randomized independently of the epidemic state conditional on
$ H_t$. After $y_{t+1}$ is observed under design $d_{t+1}$, Bayes'
theorem gives the update step
\begin{equation}
p(x_{t+1}\mid h_{t+1})
=
\frac{
p(y_{t+1}\mid x_{t+1},d_{t+1})\,
p(x_{t+1}\mid h_t)
}{
p(y_{t+1}\mid h_t,d_{t+1})
}.
\label{eq:bg-update}
\end{equation}
The normalizing constant is the design-dependent predictive probability
\begin{equation}
p(y_{t+1}\mid h_t,d_{t+1})
=
\int
p(y_{t+1}\mid x_{t+1},d_{t+1})
p(x_{t+1}\mid h_t)
\,dx_{t+1}.
\label{eq:bg-normalising-constant}
\end{equation}
The design (sample size and its allocation across strata) therefore affects the update through the observation
model, but does not otherwise change the state predictive distribution. If the design
is informative about the next state, its generating process must also be modeled.
The state distribution after conditioning on such a design satisfies
\begin{equation}
p(x_{t+1}\mid h_t,d_{t+1})
\propto
p(d_{t+1}\mid x_{t+1}, h_t)
p(x_{t+1}\mid h_t).
\label{eq:bg-informative-design}
\end{equation}

The recursion is exact, but the required integrals only have closed-form or
finite-dimensional exact solutions for models with suitable structure. The
linear Gaussian model is the principal example. Given a Gaussian initial
distribution, linear state and observation equations, and Gaussian process and
observation noise, the filtering distribution remains Gaussian and is represented
exactly by its mean and covariance \citep{kalman_new_1960,jazwinski_stochastic_1970}.
With nonlinear dynamics or non-Gaussian observations, the filtering distribution
may be skewed, bounded, discrete, or multimodal, so a mean and covariance do not
generally provide a complete representation.

Approximate Kalman-type filters address this using local or
moment-based approximations. The extended Kalman filter linearizes the transition
and observation functions around the current estimate and higher-order variants keep additional terms in the local expansion \citep{jazwinski_stochastic_1970}. The unscented Kalman filter propagates a deterministic set
of sigma points to approximate the transformed mean and covariance
\citep{julier_unscented_2004}. The ensemble Kalman filter instead propagates a
stochastic ensemble and uses its empirical mean and covariance in a Kalman-style
update \citep{evensen_sequential_1994}. These methods can perform well when the
evolving uncertainty in $x_t$ is adequately represented by its first two moments. Mechanistic epidemic models often violate these
assumptions. Their states are constrained, dynamics are nonlinear, parameters interact with latent states, and their observations often follow
discrete count distributions. Sequential Monte Carlo methods approximate the Bayesian recursion without requiring
Gaussian closure, which makes them better suited to these models
\citep{arulampalam_tutorial_2002}.

\subsection{Sequential Monte Carlo and the particle filter}
\label{sec:bg-smc}

Particle filters are sequential Monte Carlo (SMC) methods that approximate a
sequence of filtering distributions using weighted samples called particles
\citep{doucet_introduction_2001}. Let $N$ be the number of particles. If
particle $i$ at time $t$ has state $x_t^{(i)}$ and normalized weight
$w_t^{(i)}$, the particle approximation to the filtering distribution is the
discrete probability measure
\begin{equation}
\widehat p_t^N(dx_t\mid h_t)
=
\sum_{i=1}^N w_t^{(i)}\delta_{x_t^{(i)}}(dx_t),
\qquad
\sum_{i=1}^N w_t^{(i)}=1,
\label{eq:bg-particle-measure}
\end{equation}
where $\delta_{x_t^{(i)}}$ is the Dirac measure at $x_t^{(i)}$.

The general sequential importance sampling update propagates a particle from
$x_t^{(i)}$ using a proposal density
\begin{equation}
x_{t+1}^{(i)}
\sim
q_{t+1}\!\left(\,\cdot\mid x_t^{(i)},y_{t+1},d_{t+1}\right).
\label{eq:bg-smc-proposal}
\end{equation}
Before any resampling, its unnormalized weight is
\begin{equation}
\widetilde w_{t+1}^{(i)}
=
w_t^{(i)}
\frac{
p(y_{t+1}\mid x_{t+1}^{(i)},d_{t+1})
p(x_{t+1}^{(i)}\mid x_t^{(i)})
}{
q_{t+1}(x_{t+1}^{(i)}\mid x_t^{(i)},y_{t+1},d_{t+1})
},
\qquad
w_{t+1}^{(i)}
=
\frac{\widetilde w_{t+1}^{(i)}}
{\sum_{j=1}^N\widetilde w_{t+1}^{(j)}}.
\label{eq:bg-smc-weight}
\end{equation}
The bootstrap particle filter uses the state transition as the proposal. In
that case the transition terms cancel from Eq.~\eqref{eq:bg-smc-weight}, and
the incremental weight is the observation likelihood
$p(y_{t+1}\mid x_{t+1}^{(i)},d_{t+1})$.

Early instances of sequential importance sampling for nonlinear filtering
include the methods of \citet{handschin_monte_1969} and
\citet{akashi_random_1977}. Repeated importance weighting generally causes the
normalized weights to become increasingly concentrated. In practice, only a
small number of particles may then retain appreciable weight, which is known as
weight degeneracy. A commonly used diagnostic for weight degeneracy is the estimated effective particle filter sample
size
\begin{equation}
\widehat N_{\mathrm{eff},t}
=
\frac{1}{\sum_{i=1}^N\left(w_t^{(i)}\right)^2},
\qquad
1\leq \widehat N_{\mathrm{eff},t}\leq N.
\label{eq:bg-ess}
\end{equation}
This quantity measures weight concentration and can be interpreted as the
equivalent number of equally weighted particles \citep{kong_sequential_1994}.

The bootstrap filter of \citet{gordon_novel_1993} combines propagation from the
state transition, likelihood weighting, and resampling in a practical recursive
filter. Resampling draws ancestor indices according to normalized weights
and assigns equal weights to the resulting particles. Particles with large
weights are therefore more likely to be replicated, while those with small
weights are more likely to be removed. A closely related Monte Carlo filter and
smoother was developed by \citet{kitagawa_monte_1996}. Both methods  mitigate
weight degeneracy via resampling, but they can lead to reduced particle diversity when several  replicated particles are assigned to the same state, a phenomenon known as sample impoverishment
\citep{arulampalam_tutorial_2002}.

These methods were consolidated by
\citet{liu_sequential_1998} and \citet{doucet_sequential_2000}, who assembled a
general SMC framework from
proposal construction, importance weighting, resampling, Markov chain Monte
Carlo moves, and Rao--Blackwellization which
partitions the state and marginalizes the analytically tractable component so
that Monte Carlo error affects only the remainder
\citep{casella_rao-blackwellisation_1996}.

\subsection{Particle filtering for infectious-disease models}
\label{sec:bg-pf-id}

Particle filters do not require linear dynamics, Gaussian distributions, or an
analytically tractable filtering recursion. Suppose an epidemic model describes
a partially observed latent process with nonlinear compartmental dynamics,
linked to surveillance data through discrete count distributions. A particle
filter can be applied to such a model provided that particles can be simulated
from the transition or proposal distribution and the observation likelihood
can be evaluated.

Particle methods have been used for different objectives in infectious-disease
analyses. \citet{dukic_tracking_2012} used an online particle filter to update
influenza states and parameters as surveillance observations arrived and
construct real-time predictive distributions. For measles,
\citet{he_plug-and-play_2010} used sequential Monte Carlo within iterated
filtering for likelihood-based parameter estimation. Particle filters have also
provided likelihood estimates within particle Markov chain Monte Carlo methods
for reconstructing latent epidemic trajectories and time-varying transmission
parameters in influenza, Ebola, and dengue models
\citep{dureau_capturing_2013,funk_real-time_2018,cazelles_accounting_2018}.

However, static parameter learning requires additional care because
repeated resampling can progressively remove parameter diversity
\citep{kantas_particle_2015}. A second
challenge arises when a new observation lies in a region assigned little
probability by the propagated particles. The resulting likelihood may
concentrate almost all weight on a small number of particles. This problem can
be aggravated by large or precise observation vectors, high-dimensional states,
model misspecification, or insufficient process variation
\citep{snyder_obstacles_2008}. These challenges have motivated methods for dynamic parameter estimation, which give parameters artificial dynamics or kernel-smoothed rejuvenation so that diversity is replenished under resampling \citep{liu_combined_2001,kantas_particle_2015}, and for observation-informed particle propagation, which uses the incoming observation to guide ancestor selection or to move particles toward it before weighting \citep{pitt_filtering_1999,zuo_adaptive_2013}.

\subsection{Related work}
\label{sec:bg-related-work}

The assurance-on-width criterion has been used to estimate sample size of diagnostic accuracy
studies using a design prior fixed in advance \citep{wilson2022,williamson_sample_2023}. The closest surveillance
tool is SASSE \citep{smith_sasse_2025}, a Bayesian planning and teaching tool where the analyst can revise the prior inputs  between rounds, though it does not propagate
a fitted state-space model to obtain a next-round predictive distribution.

Other adaptive surveillance methods decide where, when, or what to sample under
a fixed resource level. Examples include adaptive geostatistical designs for
prevalence mapping \citep{chipeta_adaptive_2016}, Bayesian experimental design
for tick surveillance \citep{case_adapting_2024}, information-guided selection
of surveillance targets \citep{tan_information-guided_2025}, test allocation in
contact networks \citep{batlle_adaptive_2022}, and reallocation of genomic
sequencing between points of entry and communities
\citep{contreras_model-based_2023}. These methods are not specifically for deciding the sample size for the next round. \citet{perez-vieites_online_2026} come closest methodologically,
combining nested particle filters with an expected-information-gain criterion
for online design in partially observed systems, including an SIR model. Their
criterion targets learning model parameters rather than a precision
requirement on a surveillance estimand. To our knowledge, no existing method
uses a filtering predictive distribution with a precision assurance criterion
to set the minimum sample size for the next surveillance round.

\section{Adaptive sampling framework}
\label{sec:framework}

The framework proposed here can dynamically determine how many samples are needed for the next surveillance round, and where those samples should be allocated to estimate a dynamic surveillance target with prespecified precision. The target may be any time-varying estimand defined by the state-space model, such as incidence or, as in the case study below, infection-induced seroprevalence. As new data arrive, a particle filter assimilates them recursively, revises the current distribution of the latent epidemic state, and propagates that distribution forward to obtain a short-term predictive distribution for the evolving target. The Bayesian assurance rule then evaluates candidate sampling designs over the possible future datasets implied by this predictive distribution and recommends a minimum-cost admissible design for which the posterior interval is predicted to meet the target width with the prespecified assurance probability. This makes the sampling design responsive to the evolving epidemic state.

As discussed in Section~\ref{sec:bg-decision}, the assurance rule is Bayesian in its use of the posterior predictive distribution, but it is not necessarily a full expected-utility design. A  fully decision-theoretic design could instead maximize an expected utility such as Lindley's expected information gain, with sampling cost penalized directly or imposed as a budget constraint \citep{lindley1956}. We use a constraint-based criterion, following the interval-based Bayesian sample-size rules \citep{lindley1997,joseph1995,joseph1997}, because a tolerable interval width and an acceptable probability of achieving it are easier for surveillance planners to specify than a utility for information gain.

\subsection{Surveillance history and state-space model}
\label{sec:fw-statespace}

Consider a surveillance system observed over discrete rounds $t=1,2,\ldots$. The latent state $x_t$ has initial distribution $p_1(x_1)$ and evolves according to a (possibly time-varying) Markov transition model $p_t(x_{t+1}\mid x_t)$, so that, given $x_t$, the next state is independent of earlier states, observations, and designs. Before the observation in round $t$ is generated, a design $d_t$ is chosen from an admissible set $\mathcal{D}_t$. The design contains the operational quantities under the analyst's control, such as the total sample size and, if relevant, its allocation across strata. Given $(x_t,d_t)$, the observation $Y_t$ is independent of the preceding history and follows the observation model $p_t(y_t\mid x_t,d_t)$, with realized value $y_t$, which may be scalar or vector valued. For notational simplicity, we assume the state is written in coordinates under which the distributions arising in the recursion admit densities with respect to a reference measure. We write $dx_t$ for integration with respect to that measure. For discrete variables, densities are interpreted as probability mass functions and integrals as sums.

At decision time $t$, the next design $d_{t+1}$ is chosen using only the surveillance history $h_t$ of Section~\ref{sec:bg-adaptive}. We assume that the design is non-informative about the next state after conditioning on this history, so that, for every candidate $d_{t+1}\in\mathcal{D}_{t+1}$,
\begin{equation}
p(x_{t+1}\mid d_{t+1}, h_t)
=
p(x_{t+1}\mid h_t).
\label{eq:fw-noninformative-design}
\end{equation}
This is Eq.~\eqref{eq:bg-noninformative-design} for a candidate design, and it holds under the conditions given there. The surveillance estimand is $\psi_t=\varphi(x_t)$. For interval-width design, $\psi_t$ is scalar; more generally, the framework may use any estimand for which a scalar posterior precision functional has been specified.

\subsection{Sequential predictive distribution}
\label{sec:fw-predictive}

Given the filtering distribution $p(x_t\mid h_t)$, the state predictive distribution for the next round is
\begin{equation}
p(x_{t+1}\mid h_t)
=
\int p_t(x_{t+1}\mid x_t)\,p(x_t\mid h_t)\,dx_t.
\label{eq:fw-state-predictive}
\end{equation}
By Eq.~\eqref{eq:fw-noninformative-design}, this distribution is also $p(x_{t+1}\mid d_{t+1}, h_t)$ for every candidate design $d_{t+1}\in\mathcal{D}_{t+1}$. For notational convenience we write $d$ for a candidate $d_{t+1}$ from here on, and retain the subscript for the selected design $d_{t+1}^{*}$. Combining the state predictive with the observation model gives the design-dependent predictive distribution of the next observation,
\begin{equation}
m_{t+1,d}(y)
=
p(y\mid d, h_t)
=
\int p_{t+1}(y\mid x_{t+1},d)\,p(x_{t+1}\mid h_t)\,dx_{t+1}.
\label{eq:fw-predictive}
\end{equation}
For any prospective observation $y$ with $m_{t+1,d}(y)>0$, the hypothetical posterior state distribution is
\begin{equation}
p(x_{t+1}\mid y,d, h_t)
=
\frac{
p_{t+1}(y\mid x_{t+1},d)\,p(x_{t+1}\mid h_t)
}{
m_{t+1,d}(y)
},
\label{eq:fw-hypothetical-posterior}
\end{equation}
where conditioning on $y$ and $d$ abbreviates conditioning on observing $Y_{t+1}=y$ under candidate design $d$. The corresponding posterior distribution of $\psi_{t+1}=\varphi(x_{t+1})$ is induced by Eq.~\eqref{eq:fw-hypothetical-posterior}; for every set $B$ in the estimand space,
\begin{equation}
\Pr(\psi_{t+1}\in B\mid y,d, h_t)
=
\int \mathbf{1}\!\left\{\varphi(x_{t+1})\in B\right\}
p(x_{t+1}\mid y,d, h_t)\,dx_{t+1}.
\label{eq:fw-estimand-posterior}
\end{equation}
Thus, the design calculation requires the joint relationship between the future state $x_{t+1}$, estimand, and observation, not only the marginal predictive distribution of the estimand. When the filtering recursion has no closed form, a particle filter can be used to approximate $p(x_t\mid h_t)$ by a weighted ensemble, propagate that ensemble through the transition model, and combine the propagated particles with the observation model to approximate Eqs.~\eqref{eq:fw-predictive}--\eqref{eq:fw-estimand-posterior}. In the scalar sample-size case, the full candidate design $d$ reduces to a sample size $n$ and the predictive distribution may be written $m_{t+1,n}$. More generally, a scalar total $n$ may index a prespecified vector-valued design $d=g_{t+1}(n; h_t)$, such as when samples are  allocated across strata.

\subsection{Assurance-based design rule}
\label{sec:fw-design}

For a scalar estimand, let $I_{\tau,t+1}(y,d; h_t)$ be a prespecified rule for constructing a $100(1-\tau)\%$ posterior credible interval from Eq.~\eqref{eq:fw-estimand-posterior} (e.g., an equal-tailed or highest-posterior-density interval), and let
\begin{equation}
W_{\tau,t+1}(y,d; h_t)
=
\operatorname{width}\!\left\{I_{\tau,t+1}(y,d; h_t)\right\}.
\label{eq:fw-width-functional}
\end{equation}
For a non-scalar estimand, $W_{\tau,t+1}$ may instead be any prespecified nonnegative scalar measure of posterior precision, such as credible-region volume or maximum marginal interval width. Given a target $w^*$ and a width functional that is measurable in $y$, the assurance attained by design $d$ is the history-conditional predictive probability
\begin{equation}
A_{t+1}(d\mid h_t)
=
\int
\mathbf{1}\!\left\{
W_{\tau,t+1}(y,d; h_t)\le w^*
\right\}
m_{t+1,d}(y)\,dy.
\label{eq:fw-assurance}
\end{equation}
Given an assurance level $q\in(0,1)$, define the feasible set
\begin{equation}
\mathcal{D}_{t+1}^{\mathrm{feas}}( h_t)
=
\left\{
d\in\mathcal{D}_{t+1}:
A_{t+1}(d\mid h_t)\ge q
\right\}.
\label{eq:fw-feasible-set}
\end{equation}
Provided this set is nonempty and the minimum is attained, a recommended next-round design is any
\begin{equation}
d_{t+1}^{*}
\in
\arg\min_{d\in\mathcal{D}_{t+1}^{\mathrm{feas}}( h_t)} C(d),
\label{eq:fw-design}
\end{equation}
where $C(d)$ is the cost of data collection under each candidate design. Equivalently,
\begin{equation}
d_{t+1}^{*}
\in
\arg\min_{d\in\mathcal{D}_{t+1}} C(d)
\quad\text{subject to}\quad
\Pr_{Y\sim m_{t+1,d}}\!\left\{
W_{\tau,t+1}(Y,d; h_t)\le w^*
\right\}\ge q,
\label{eq:fw-design-constrained}
\end{equation}
where a deterministic tie-breaking rule may be imposed if a unique recommendation is required. If the feasible set is empty, the design problem is infeasible over $\mathcal{D}_{t+1}$. A surveillance system that must sample in every round can then take the largest admissible design and accept an interval wider than $w^*$, which keeps the recursion running.

Assurance need not be monotone in the total sample size $n$, so a rule that selects the first sample size at which assurance crosses the target may select a size that satisfies the constraint while a larger size does not. We therefore use the sustained assurance

\begin{equation}
A_{t+1}^{\mathrm{sus}}(n\mid h_t)
=
\inf_{\substack{m\in\mathcal{N}_{t+1}\\m\ge n}}
A_{t+1}(m\mid h_t),
\label{eq:fw-sustained-assurance}
\end{equation}

which is nondecreasing in $n$. When cost is strictly increasing in $n$, the selected sample size is

\begin{equation}
n_{t+1}^{\mathrm{ASSUR}}
=
\min\left\{
n\in\mathcal{N}_{t+1}:
A_{t+1}^{\mathrm{sus}}(n\mid h_t)\ge q
\right\},
\label{eq:fw-scalar}
\end{equation}

provided this set is nonempty.

The criterion values data only for the precision the model expects to gain. It gives no value to data as a check on the model, so a minimum sample size must be set through the admissible set $\mathcal{N}_{t+1}$, for example with a floor $n_{\min}$. The credibility level, interval rule, admissible designs, and cost function are chosen by the analyst.

Some designs require allocating a total sample across strata or sites, so that
the candidate design is a vector $d=(n_1,\ldots,n_S)$. Although
Eq.~\eqref{eq:fw-design} can be applied directly to all admissible allocation
vectors, the allocation and induced outcome spaces grow rapidly with $n$. For the
ten-stratum case study of Section~\ref{sec:casestudy}, a total of 130 samples can be
allocated in about $4\times10^{13}$ ways, with about $3\times10^{38}$
allocation--outcome pairs under the four-cell observation model. A prespecified
allocation rule can instead map each admissible total $n\in\mathcal{N}_{t+1}$
to a vector
\[
g_{t+1}(n; h_t)
=
\bigl(n_1(n),\ldots,n_S(n)\bigr).
\]
Once any tie-breaking rule is fixed, this defines the scalar-indexed design
family
\[
\mathcal{D}_{t+1}^{g}( h_t)
=
\left\{
g_{t+1}(n; h_t):
n\in\mathcal{N}_{t+1}
\right\}.
\]
Define the assurance induced by this family as
$A_{t+1}^{g}(n\mid h_t)
=A_{t+1}\{g_{t+1}(n; h_t)\mid h_t\}$.
The sustained-assurance construction in
Eq.~\eqref{eq:fw-sustained-assurance} then applies to
$A_{t+1}^{g}$ over $\mathcal{N}_{t+1}$. This yields a formal assurance
statement for the selected allocation vector, but minimum cost is asserted only
within $\mathcal{D}_{t+1}^{g}$, not over all possible allocation vectors. 

Algorithm~\ref{alg:framework} summarizes the adaptive sampling loop.

\begin{algorithm}[H]
\caption{Adaptive sampling loop}
\label{alg:framework}
\KwIn{Process and observation models, initial filtering distribution, interval rule, admissible designs and costs, $w^*$, $q$}
\For{each decision time $t$}{
  \textbf{Filtering stage}\\
  Propagate $p(x_t\mid h_t)$ to $p(x_{t+1}\mid h_t)$ (Eq.~\eqref{eq:fw-state-predictive})\;
  \textbf{Design stage}\\
  For each $d\in\mathcal{D}_{t+1}$, compute $A_{t+1}(d\mid h_t)$ (Eq.~\eqref{eq:fw-assurance}), or its sustained version (Eq.~\eqref{eq:fw-sustained-assurance}) for scalar-indexed designs\;
  \eIf{some candidate attains $q$}{
    Set $d_{t+1}^{*}$ to the cheapest such candidate\;
  }{
    Set $d_{t+1}^{*}$ to the largest admissible design and record the round as infeasible\;
    }
  \textbf{Sampling and update}\\
  Observe $y_{t+1}$ under $d_{t+1}^{*}$ and update to $p(x_{t+1}\mid h_{t+1})$\;
}
\end{algorithm}

\section{Simulation studies}
\label{sec:simulation}

We evaluated the filtering and design loop in two simulations. The
first simulation checks that the loop behaves as intended when the transition model is correctly specified, and the second assesses its behavior when an epidemiological change leads to a misspecified transition model. Specifically, we assess whether sampling effort is temporarily
increased when observations are surprising and the model
is learning the new parameters.

\subsection{Design response under an anticipated change in uncertainty}
\label{sec:sim-known}

In this simulation study, the latent prevalence follows a random walk on the logit scale,
moving by a random increment each round. The variance of that increment, or
``innovation variance'' sets how far the prevalence can move between rounds and
therefore how uncertain its next value is. Here, we simulate increasing the innovation variance by a known amount in two prespecified rounds in both the data-generating process and in the filter.

This is analogous to an anticipated shift in dynamics, such as a newly detected variant or change to physical distancing policies during the COVID-19 pandemic (with the caveat that in real-world applications, analysts would be unlikely to adjust the innovation variance of the filter in a way that perfectly matches the magnitude and timing of the change in the data-generating process). Our simulation examines filter tracking,
interval coverage, the response of the selected total to predictive
uncertainty, agreement between realized attainment of the width target and
model-based assurance, and sampling effort.

\paragraph{Design.}
The latent state was a two-dimensional prevalence-logit vector
$z_t=(z_{A,t},z_{B,t})^\top$. Its components were transformed to
stratum-specific prevalences by
$p_{s,t}=\operatorname{logit}^{-1}(z_{s,t})$. The state followed a correlated
Gaussian random walk,
\begin{equation}
z_0\sim\mathcal N(\mu_0,\Sigma_0),
\qquad
z_t\mid z_{t-1}
\sim
\mathcal N\!\left(z_{t-1},a_t^2\Sigma\right),
\label{eq:sim-state}
\end{equation}
over 12 surveillance rounds. The innovation standard deviations were
multiplied by three in rounds 6 and 7, so the innovation covariance increased
ninefold during those transitions.

Conditional on the latent state, the observations in the two strata were
independent binomial counts,
\begin{equation}
Y_{s,t}\mid z_t,n_{s,t}
\sim
\operatorname{Binomial}(n_{s,t},p_{s,t}),
\qquad s\in\{A,B\}.
\label{eq:sim-observation}
\end{equation}
The scalar surveillance estimand was the population-weighted prevalence
$\psi_t=0.6p_{A,t}+0.4p_{B,t}$. A bootstrap particle filter with 1000
particles approximated the filtering and one-step predictive distributions,
with systematic resampling when the effective filter sample size fell below 500.
The same model generated the latent trajectories and propagated the
particles.

Candidate total sample sizes were
$\mathcal N=\{20,40,\ldots,240\}$. Each candidate allocation began with five
observations per stratum. Remaining observations were assigned one at a time
to the stratum with the largest reduction in a working approximation to the
sampling variance of $\psi_t$. This greedy rule constructed a deterministic
scalar-indexed allocation family. Assurance was then evaluated for each
resulting allocation under the full joint particle predictive distribution,
not under the working variance approximation.

The design target was the width of a 90\% equal-tailed posterior interval for
$\psi_t$. The target width was $w^*=0.10$ and the assurance requirement was
$q=0.90$. Design-stage assurance used 1000 predictive Monte Carlo draws. After selection, an independent set of 5000 draws re-estimated the assurance of the selected design without altering the choice.
Candidate totals shared particle ancestors and nested Bernoulli streams so
that comparisons used common random numbers
\citep{glasserman_guidelines_1992}, while each candidate retained its correct
predictive distribution.
Selection was based on exact one-sided
Clopper--Pearson lower bounds with 95\% Bonferroni familywise confidence over
the candidate grid.

We simulated 100 latent trajectories, each with its own observation and
particle-filter random streams. Performance measures
were calculated within each replicate and then averaged across replicates (details in Appendix Section~\ref{app:simulation-details}).

\paragraph{Results.}
The particle approximation
tracked the simulated estimand under the correctly specified model. Under the adaptive sampling policy, mean
one-step 90\% predictive interval coverage was 0.903 (95\% Monte Carlo
interval of 0.882--0.925). Mean posterior interval coverage was 0.913
with an interval of 0.889--0.937, and the mean replicate-level posterior RMSE
was 0.025 with an interval of 0.023--0.026. The policy used a mean of 741.2 observations over the 12 rounds (95\%
Monte Carlo interval of 701.8--780.6), and its mean posterior width was 0.088, below the precision target of 0.10.

\begin{figure}[H]
    \centering
    \includegraphics[width=0.98\linewidth]{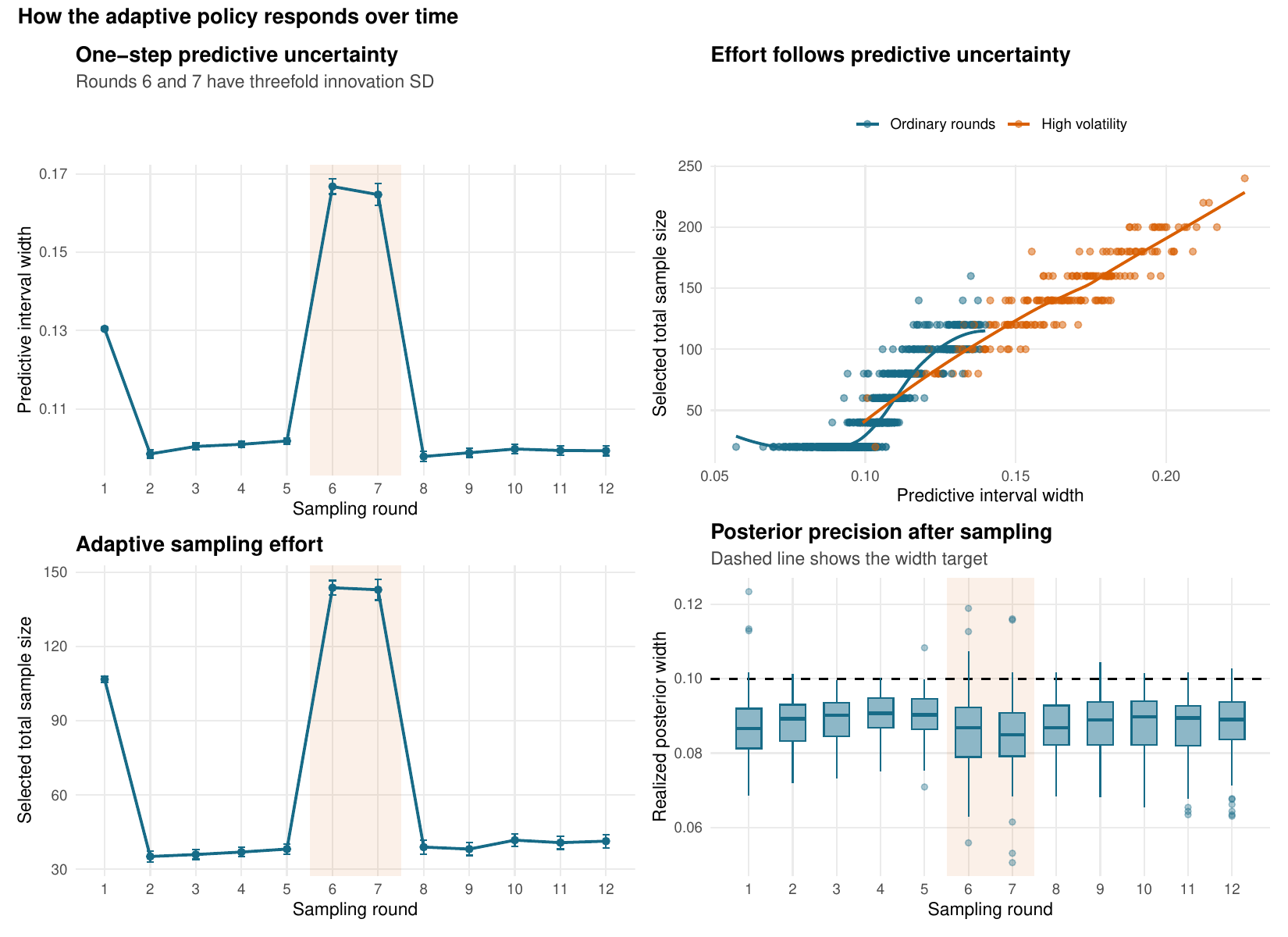}
    \caption{Response of the adaptive policy in the correctly specified simulation. Upper-left: mean one-step predictive interval width by round and its Monte Carlo standard error. Lower-left: the corresponding selected total. Shading marks high-volatility rounds. Upper-right: predictive width and selected size. Lower-right: realized posterior interval widths (target width indicated by the dashed line).}
    \label{fig:simulation-dynamic-response}
\end{figure}

Sampling effort responded to the change in process uncertainty. The mean
selected total was 45.4 in ordinary rounds and 143.4 in
the high-volatility rounds, with respective 95\% Monte Carlo intervals of
42.5--48.4 and 137.1--149.7. Within trajectories, the mean Spearman
correlation between one-step predictive width and selected total was 0.864
with an interval of 0.848--0.881. Figure~\ref{fig:simulation-dynamic-response}
shows the corresponding increase in predictive width and sampling effort in
rounds 6 and 7. Posterior widths remained concentrated near or below the
target after sampling.

The numerical assurance calculation was conservative and well calibrated.
Mean independently validated assurance was 0.974 with a 95\% Monte Carlo
interval of 0.973--0.976 and the target width was realized in 0.970 of
adaptive decisions with an interval of 0.961--0.979. Their replicate-level
difference was $-0.004$ with an interval of $-0.013$--$0.005$. The mean
sustained simultaneous lower bound at the selected designs was 0.960, and no
selected design had independently validated assurance below 0.90. The mean
absolute difference between design-stage and validation assurance estimates
was 0.004.

\begin{figure}[H]
    \centering
    \includegraphics[width=0.98\linewidth]{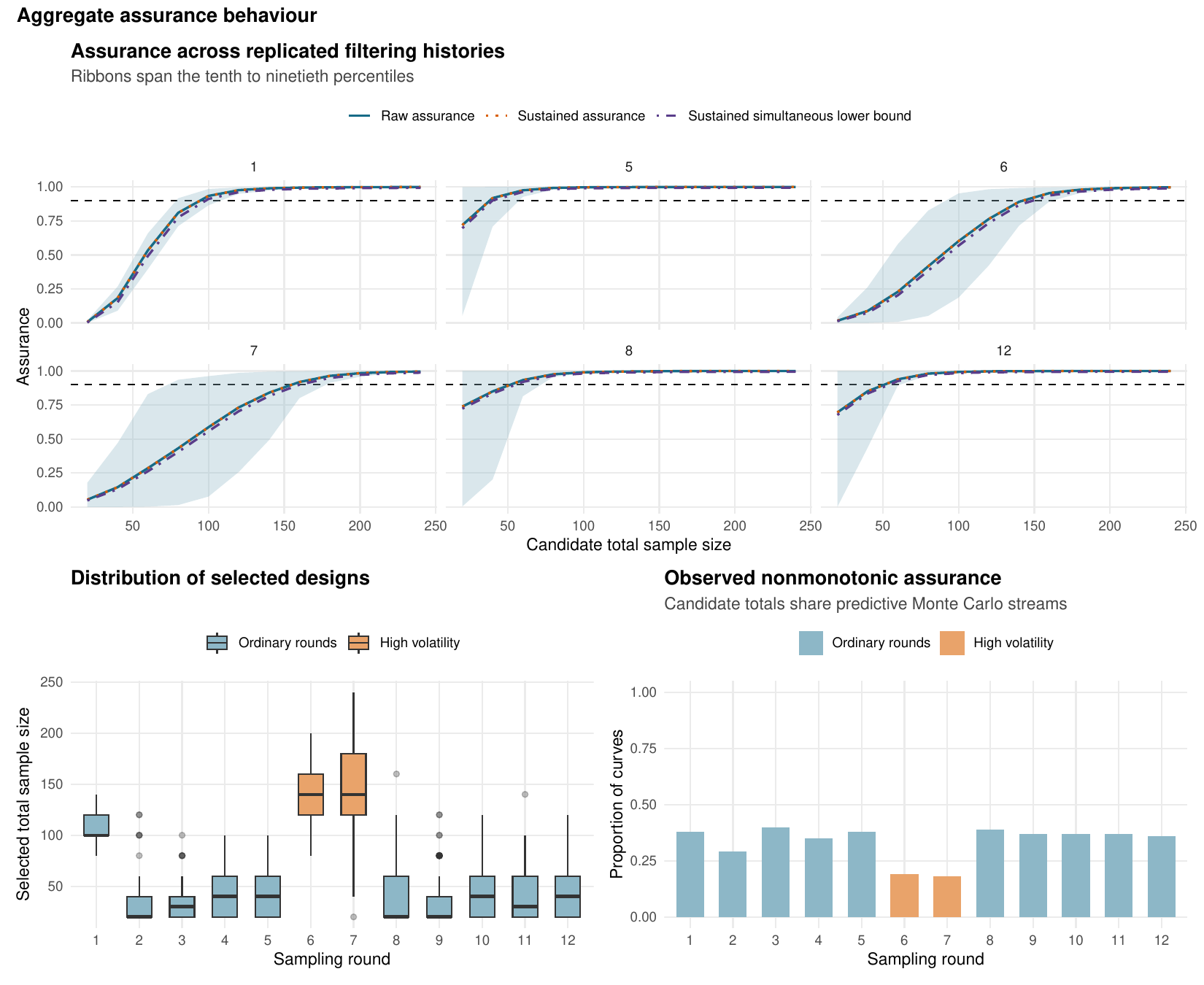}
    \caption{Raw assurance, sustained assurance and the sustained lower bound over the candidate grid (upper), and selected totals and the share of nonmonotone raw curves by round (lower).}
    \label{fig:appendix-simulation-assurance}
\end{figure}

\begin{figure}[H]
    \centering
    \includegraphics[width=0.98\linewidth]{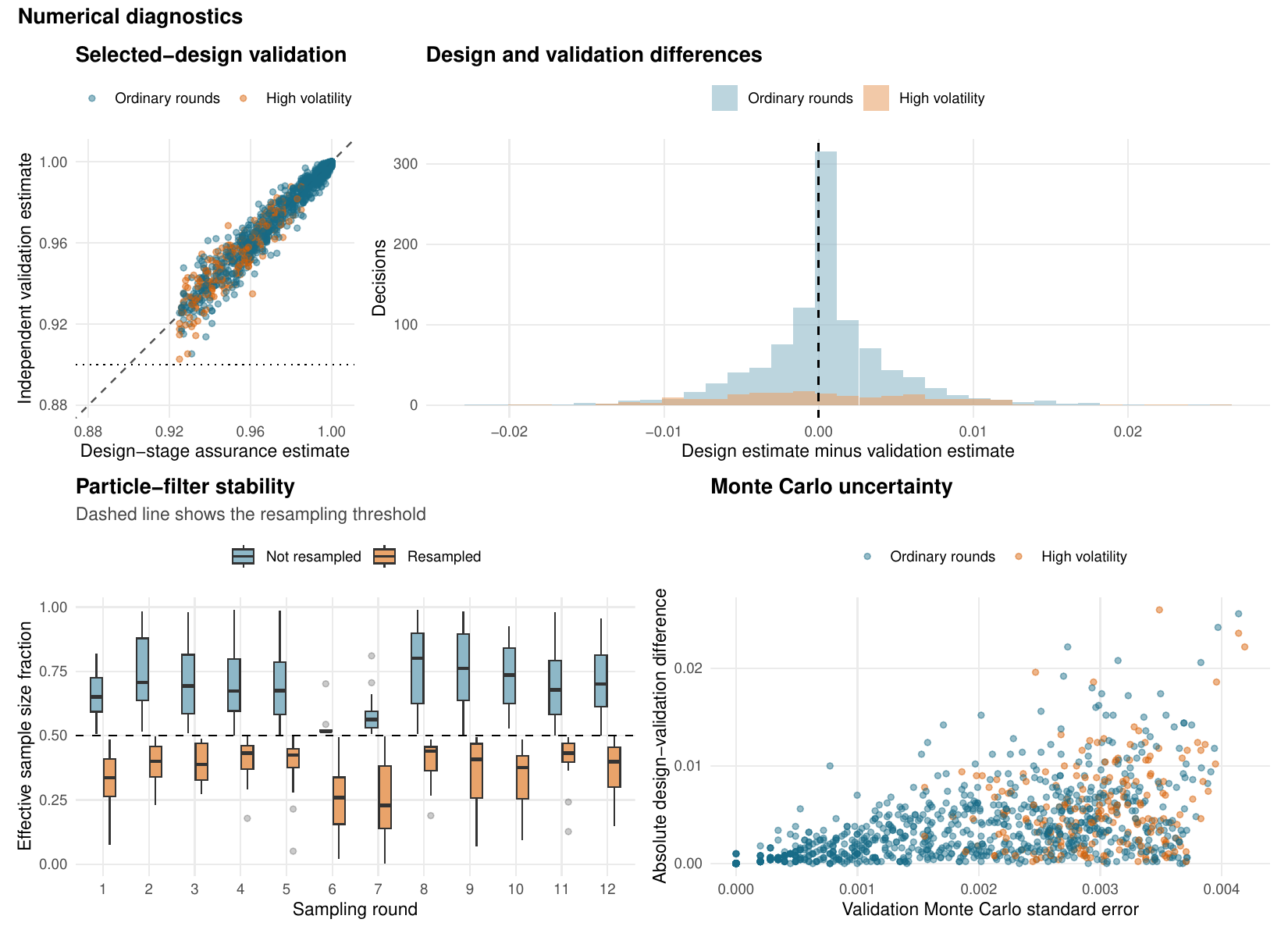}
    \caption{Design-stage against validated assurance, their differences, the effective number of particles by round, and validation Monte Carlo error against the design-validation difference.}
    \label{fig:appendix-simulation-numerical}
\end{figure}

Detailed component summaries and numerical
diagnostics are given in Appendix Table~\ref{tab:simple_simulation_summary}
and Appendix Figures~\ref{fig:appendix-simulation-assurance}
and~\ref{fig:appendix-simulation-numerical}.

\subsection{Design response under an unanticipated change}
\label{sec:sim-unannounced}

In this simulation, repeated serological surveys track an epidemic that follows a
compartmental transmission model. At one round the transmission rate is multiplied by a
fixed factor, representing an event similar to introduction of a more transmissible variant.
The filter's transition model
contains no such jump, so the change reaches it only through observations the
model did not predict. The study measures whether the filter recognizes the
change from the resulting prediction errors, whether the design raises
sampling effort while the transmission rate is being relearned, and whether
the assurance calculation remains accurate.

\paragraph{Design.}
Two strata, $A$ and $B$, carry population weights
$(\omega_A,\omega_B)=(0.6,0.4)$. Within stratum $s$ the compartment
proportions $(S_{s,t},I_{s,t},R_{s,t})$ sum to one and evolve according to
\begin{equation}
\frac{dS_{s,t}}{dt}=-\lambda_{s,t}S_{s,t},
\qquad
\frac{dI_{s,t}}{dt}=\lambda_{s,t}S_{s,t}-\gamma I_{s,t},
\qquad
\frac{dR_{s,t}}{dt}=\gamma I_{s,t},
\label{eq:sir-system}
\end{equation}
with force of infection
$\lambda_{s,t}=\beta_t\sum_{s'\in\{A,B\}}M_{s,s'}I_{s',t}$
over 34 rounds. Only the recovered compartment is observed through serology,
so the estimand is the population-weighted seroprevalence
$\psi_t=\omega_AR_{A,t}+\omega_BR_{B,t}$. The transmission rate followed a geometric random walk
carrying a single unanticipated multiplicative jump,
\begin{equation}
\beta_t
=
\beta_{t-1}
\exp(\varepsilon_t)
\kappa^{\mathbf 1\{t=t_{\mathrm v}\}},
\qquad
\varepsilon_t\sim\mathcal N(0,\sigma_{\mathrm{true}}^2),
\label{eq:sir-beta}
\end{equation}
with $\sigma_{\mathrm{true}}=0.04$, jump size $\kappa=1.55$, and jump round
$t_{\mathrm v}=12$. The initial basic reproduction number was 1.25 and the
jump raised it to approximately 1.94. The jump occurs in a single round. We summarize results before the jump
(rounds 1 to 11), in a transition window while the filter relearns the
transmission rate, and after that window. The window starts at the jump round
and lasts for the mean transmission-rate recovery lag, the number of rounds
until the posterior mean of $\beta$ is within 15\% of the truth for two
consecutive rounds. This lag averaged 5.2 rounds (reported below), so the
window covers rounds 12 to 16 and the period after it covers rounds 17 to 34.
Conditional on the state, the stratum counts were independent
binomial variables as in Eq.~\eqref{eq:sim-observation}.

The filter also carries
a larger random-walk standard deviation than the data-generating process. The design components were those
of Section~\ref{sec:fw-design}, with a 90\% equal-tailed interval for
$\psi_t$, $w^*=0.05$, and $q=0.90$. The study
used 100 independent replicates. Appendix Section~\ref{app:sir-variant} provides
the contact matrix, integration scheme, particle initialization, candidate
grid, Monte Carlo settings, and the definitions of the recovery measures.

\paragraph{Reference rule.}
For a fixed sample size, the interval width for a proportion is largest near
0.5, and $\psi_t$ passes through that region during the wave, so the selected
total would rise even if predictive uncertainty were constant. We therefore
compare it with a reference computed at the true $\psi_t$ without the filter.
Let
\begin{equation}
P_t^{\mathrm{ref}}(n)
=
\Pr\!\left\{W_{\tau}^{\mathrm B}(Y,n)\le w^*\right\},
\qquad
Y\sim\operatorname{Binomial}(n,\psi_t),
\label{eq:sir-reference-assurance}
\end{equation}
where $W_{\tau}^{\mathrm B}(y,n)$ is the width of the equal-tailed
$100(1-\tau)\%$ posterior interval under the Jeffreys prior, and let
\begin{equation}
n_t^{\mathrm{ref}}
=
\min\left\{n:\min_{n'\ge n}P_t^{\mathrm{ref}}(n')\ge q\right\},
\label{eq:sir-reference}
\end{equation}
with $n$ and $n'$ ranging over the positive integers. The reference uses the
unobserved estimand and no information from earlier rounds. The ratio
$n_t/n_t^{\mathrm{ref}}$ approximately measures the effort saved by the
information the filter carries. It also reflects rounding of $n_t$ to the
candidate grid and the stratified allocation, and we report it for rounds in which
$n_t$ exceeded the grid floor.

\begin{figure}[H]
    \centering
    \includegraphics[width=0.98\linewidth]{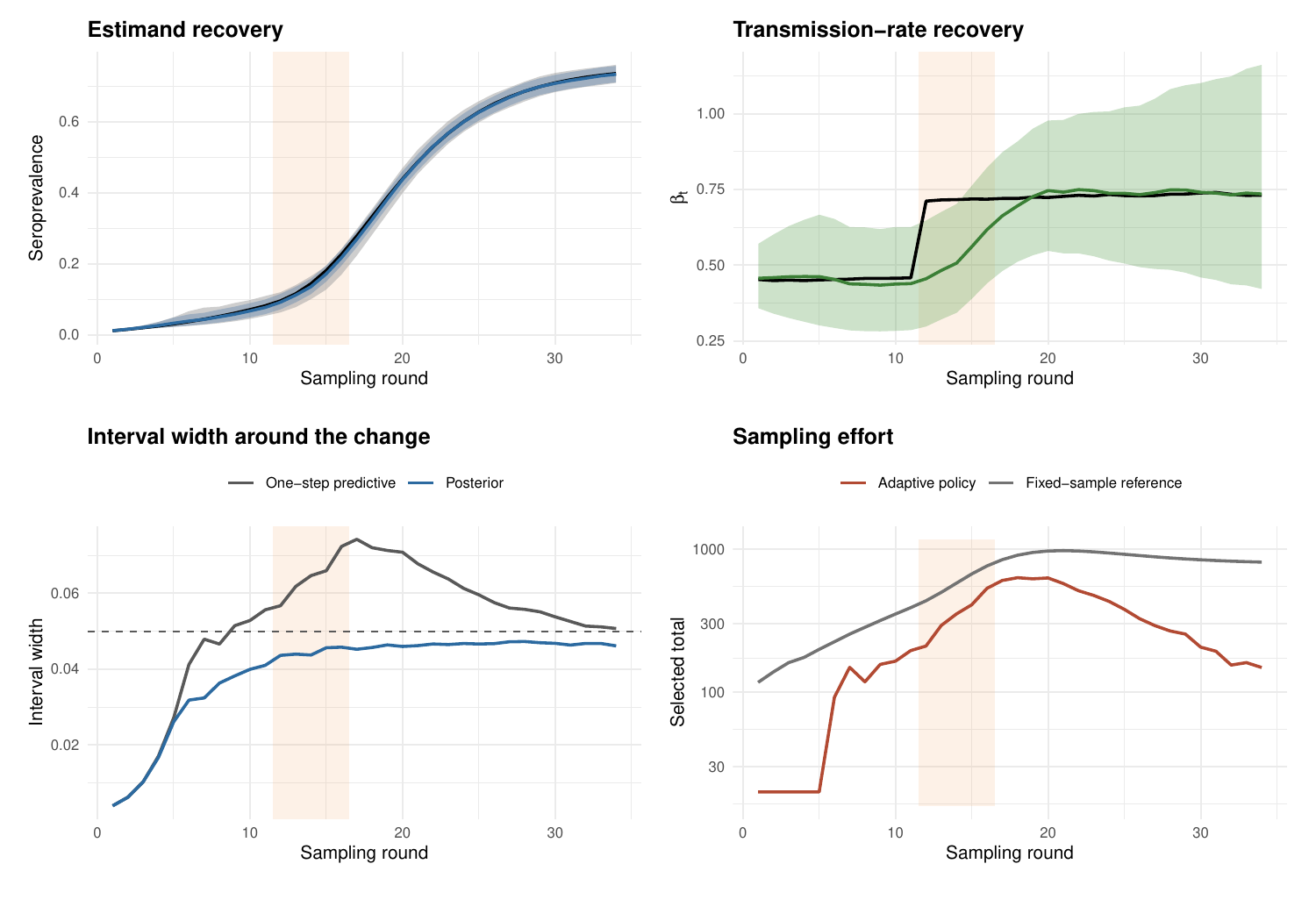}
    \caption{Stratified SIR simulation with an unanticipated change in transmission. Shading marks the transition window. Upper-left: true estimand (black line) with the one-step predictive (black shading) and posterior ribbons (blue shading). Upper-right: true transmission rate (black line) and the posterior mean and 90\% interval (green line and shading). Lower-left: mean interval widths with the target width indicated by the dashed line. Lower-right: selected total against the variance-matched reference rule of Eq.~\eqref{eq:sir-reference} on a logarithmic scale.}
    \label{fig:sir-variant-dynamics}
\end{figure}

\paragraph{Results.}
The mean selected
sample size per round rose from 88.6 before the jump to 359.2 in the transition window. Measured against the reference rule of
Eq.~\eqref{eq:sir-reference}, relative effort was 0.629 with a 95\% Monte
Carlo interval of 0.599--0.659 before the jump and 0.615 with an interval of
0.596--0.633 in the transition window, and then fell to 0.436 with an interval of
0.419--0.453 afterwards, a decline of 26.1\% with an interval of 21.6--30.6.
Sampling effort relative to the reference stayed near its pre-jump level
through the transition window, while the transmission rate was poorly estimated, and
fell after the filter had learned the new rate
(Figure~\ref{fig:sir-variant-dynamics}).

The mean
trajectory recovery lag was 2.76 rounds with an interval of 2.34--3.18, and
the mean transmission-rate recovery lag was 5.22 rounds with an interval of
4.84--5.60, which sets the length of the transition window. Accumulating observations constrained the estimand directly, whereas
the transmission rate was identified only through the shape of the trajectory,
and sampling effort remained elevated over the longer interval during which
that rate is still poorly determined. Interval coverage for the estimand and
for the transmission rate both degraded in the transition window and
returned to at least the nominal 0.90 after the window without intervention; Appendix
Section~\ref{app:sir-variant} reports these by period.

\begin{figure}[H]
    \centering
    \includegraphics[width=0.98\linewidth]{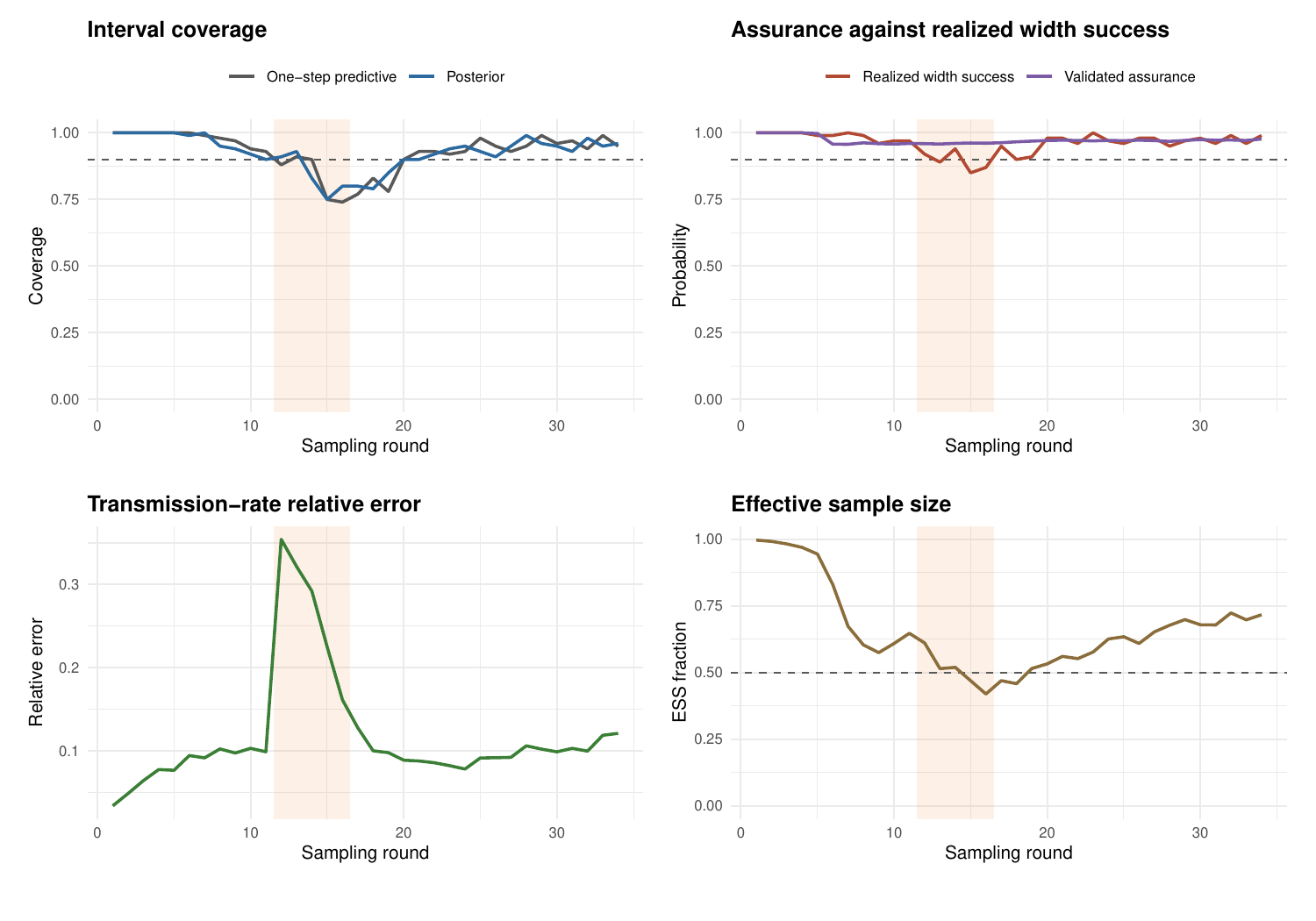}
    \caption{Diagnostics in the unanticipated-change simulation. Shading marks the transition window. Panels show interval coverage, validated assurance against realized width success, the relative error of the posterior mean transmission rate, and the effective-particle fraction.}
    \label{fig:appendix-sir-variant-calibration}
\end{figure}

The assurance calculation was less accurate compared to the first simulation. Outside the transition window it was
accurate or conservative, with realized width success of 0.988 against
validated assurance of 0.977 before the jump and 0.965 against 0.971
afterwards. In the transition window, validated assurance remained at 0.960 with
an interval of 0.958--0.962 while realized width success fell to 0.894 with an
interval of 0.869--0.919. The assurance therefore overstated the achieved
precision by about 0.066. The point estimate of realized success was below the
requested $q=0.90$, but its interval includes 0.90. The effective-sample-size fraction averaged 0.659 over all rounds and fell to 0.42 at round 16, the last round of the transition window.
Filter diagnostics signal in real time that the predictive distribution has become unreliable after the jump. Appendix
Figure~\ref{fig:appendix-sir-variant-calibration} shows these quantities by
round.

\section{Case study: SARS-CoV-2 blood donor serosurveillance}
\label{sec:casestudy}

We instantiate the framework of Section~\ref{sec:framework} for SARS-CoV-2 serosurveillance among Canadian blood donors. The serosurveillance program assessed presence of antibodies against SARS-CoV-2 spike protein (anti-S), which result from natural infection or vaccination, and antibodies against SARS-CoV-2 nucleocapsid (anti-N) which result from natural infection only. We define the surveillance estimand $\psi_t=\varphi(x_t)$ as the population-weighted anti-nucleocapsid (anti-N) seroprevalence. Within a region, the latent state $x_t$ is an age--sex-stratified stochastic compartmental epidemic model and the filtering distribution is approximated by a particle filter. The particle approximation sets the next round's total sample size by assurance on the regional estimand, and a prespecified greedy rule then allocates that total across age--sex strata.

\subsection{Data and surveillance setting}
\label{sec:cs-data}

The SARS-CoV-2 serosurveillance study was approved by the Research Ethics Board of Canadian Blood Services. This secondary analysis was approved by the Faculty of Medicine \& Health Sciences Research Ethics Board Office, McGill University, Montreal, Canada (REB number 22-03-077).

During the COVID-19 serosurveillance program, Canadian Blood Services tested residual EDTA plasma retention samples from blood donors, changing assay platform as the pandemic evolved. In the early phase of surveillance, samples were screened for IgG antibodies against the SARS-CoV-2 nucleocapsid (N) protein using the Abbott Architect SARS-CoV-2 IgG chemiluminescent microparticle immunoassay (CMIA) performed on Abbott Architect immunoassay analyzers, which allowed detection of antibodies generated following natural infection \citep{saeed_seroprevalence_2021}. That assay reports anti-N alone. As vaccination began in 2021, the program adopted Roche Elecsys dual-assay testing to distinguish infection- from vaccine-induced immunity \citep{obrien_omicron_2022}.

The source data span May 2020 through November 2023 and comprise 1,039,298 specimen records from 448,720 unique blood donors. Abbott testing spans 9 May 2020 to 27 January 2021 and Roche testing spans 14 December 2020 to 30 November 2023, with 35,421 specimens run on both platforms during the transition months of December 2020 and January 2021 and Roche alone from February 2021 onward.

The filter runs over the full series from May 2020 and design decisions are restricted to the dual-assay era. Weeks 1 to 42 form a burn-in period in which the filter assimilates the observed data at their realized volumes. The first decision round is the surveillance week containing 1 March 2021, week 43 of the series, and the last is week 186, giving 144 decision rounds in each of four regions (576 region-round combinations). The series is not continuous. In each region, 17 to 20 weeks between late March and mid-December 2021 contain no specimens, which leaves 124 to 127 decision rounds per region and 502 region-rounds with specimens. Replay results are reported over these 502 region-rounds. 

We excluded 323 records from donors residing in Quebec or the territories and 3,408 records from donors younger than 18. In the decision window we also excluded 586 specimens that lacked an interpreted result for one of the two Roche assays, 533 without anti-S and 53 without anti-N. The analytic sample comprises 1,034,981 specimens from 446,157 donors in nine provinces, with 163,923 specimens in the burn-in and 871,058 in the decision window (Table~\ref{tab:regional_data_overview}).

For modeling, we aggregated the data into weekly time steps $t=1,\ldots,T$, with $T=186$. For each stratum $s=1,\ldots,S$ and week $t$ we recorded the joint antibody outcome of every specimen tested. In the dual-assay period a specimen yields one of four outcomes, and we write $Y_{s,t}^{ab}$ for the number of specimens with observed anti-N status $a\in\{0,1\}$ and observed anti-S status $b\in\{0,1\}$, with $n_{s,t}=\sum_{a,b}Y_{s,t}^{ab}$. In the Abbott era only anti-N is observed, and a specimen contributes to the anti-N count pair
\begin{align}
(n_{s,t,N},\; y_{s,t,N}),
\end{align}
where $n_{s,t,N}$ is the number tested and $y_{s,t,N}$ the number seropositive for anti-N. The observed anti-N proportion is $\tilde p_{s,t,N}=y_{s,t,N}/n_{s,t,N}$ in the Abbott testing phase and $\tilde p_{s,t,N}=(Y_{s,t}^{10}+Y_{s,t}^{11})/n_{s,t}$ in the dual-assay phase.

Four regional models were fitted independently, for Atlantic Canada, the Prairies, British Columbia, and Ontario. The Atlantic and Prairie models pooled the provinces belonging to their respective regions. Within each regional model, the strata were the ten combinations of sex and age group (18--26, 27--36, 37--46, 47--56, and 57+ years). Each stratum was assigned a regional population size $N_s$ from the Statistics Canada population estimates for July 1, 2021 \citep{statcan_population_estimates}, obtained by summing single-year-of-age estimates over the stratum's age band and over the provinces in the region. The within-region population weight is $\omega_s=N_s/\sum_{r=1}^S N_r$.

\subsection{Case-study state-space model}
\label{sec:cs-statespace}

We specify the state and transition kernel of
Section~\ref{sec:fw-statespace} using an age--sex-stratified transmission
model with time-varying parameters \citep{hethcote_mathematics_2000,
dureau_capturing_2013}. Let $u_t$ denote the continuous-time boundary of
weekly round $t$. The latent state at that boundary is
\[
x_t=(z_t,\theta_t).
\]

Within each stratum $s=1,\ldots,S$, individuals occupy six compartments,
namely uninfected and unvaccinated ($UI_{UV,s}$), infected and unvaccinated
($I_{UV,s}$), seropositive and unvaccinated ($R_{UV,s}$), uninfected and
vaccinated ($UI_{V,s}$), infected and vaccinated ($I_{V,s}$), and
seropositive and vaccinated ($R_{V,s}$). We use the five free coordinates
\[
z_{s,t}
=
\left(
UI_{UV,s,t},
I_{UV,s,t},
R_{UV,s,t},
UI_{V,s,t},
I_{V,s,t}
\right)^\top,
\qquad
R_{V,s,t}=N_s-\mathbf 1^\top z_{s,t}.
\]
Thus
\[
z_t=(z_{1,t}^\top,\ldots,z_{S,t}^\top)^\top
\in\mathcal Z
=
\prod_{s=1}^S
\left\{
z\in[0,N_s]^5:
\mathbf 1^\top z\le N_s
\right\}.
\]

The dynamic parameter vector is
\begin{equation}
\theta_t
=
\left(
\beta_t,
\nu_t,
\{\eta_{s,t}\}_{s=1}^S,
\{g_{s,t}\}_{s=1}^S
\right)^\top
\in\Theta=\prod_{j=1}^J[a_j,b_j],
\qquad
J=2S+2.
\label{eq:cs-parameter-state}
\end{equation}
Here $\beta_t$ is the shared transmission rate, $\nu_t$ is the latent
vaccination rate, $\eta_{s,t}$ modifies susceptibility to infection, and
$g_{s,t}$ modifies the transition out of infection. Both modifiers are shared
by the unvaccinated and vaccinated compartments of stratum $s$. The recovery rate
$\gamma>0$ and nonnegative contact matrix $M$ are fixed. The bounds
$0\le a_j<b_j$ are listed in Appendix Table~\ref{tab:pf_config}.

Let $t_{\mathrm{vac}}$ be the first round in which the model allows
vaccination, and define the effective vaccination rate
\[
\bar\nu_t
=
\mathbf 1\{t\ge t_{\mathrm{vac}}\}\nu_t.
\]
We set $t_{\mathrm{vac}}$ to week 43, the week containing 1 March 2021 and
the first decision round. Vaccination in Canada began on 14 December 2020,
but anti-S, the only observation that separates vaccinated from
unvaccinated individuals, enters the filter only from week 43
(Section~\ref{sec:cs-filter}). Holding $\bar\nu_t$ at zero through the
burn-in keeps the vaccination rate from drifting with no data to inform
it. The model therefore enters the decision window with no vaccinated
individuals and learns early coverage from the anti-S results. The initial state has
a prespecified law
\begin{equation}
x_1\sim\Pi_1,
\qquad
\Pi_1(\mathcal Z\times\Theta)=1.
\label{eq:cs-initial-law}
\end{equation}

During $(u_t,u_{t+1}]$, the newly drawn parameters $\theta_{t+1}$ are held
fixed and the compartments satisfy
\begin{subequations}
\label{eq:cs-ode}
\begin{align}
\frac{dUI_{UV,s}}{du}
&=
-\lambda_{s}UI_{UV,s}
-\bar\nu_{t+1}UI_{UV,s},
\\
\frac{dI_{UV,s}}{du}
&=
\lambda_{s}UI_{UV,s}
-\gamma g_{s,t+1}I_{UV,s}
-\bar\nu_{t+1}I_{UV,s},
\\
\frac{dR_{UV,s}}{du}
&=
\gamma g_{s,t+1}I_{UV,s}
-\bar\nu_{t+1}R_{UV,s},
\\
\frac{dUI_{V,s}}{du}
&=
\bar\nu_{t+1}UI_{UV,s}
-\lambda_{s}UI_{V,s},
\\
\frac{dI_{V,s}}{du}
&=
\bar\nu_{t+1}I_{UV,s}
+\lambda_{s}UI_{V,s}
-\gamma g_{s,t+1}I_{V,s},
\\
\frac{dR_{V,s}}{du}
&=
\bar\nu_{t+1}R_{UV,s}
+\gamma g_{s,t+1}I_{V,s},
\end{align}
\end{subequations}
where all compartments on the right are evaluated at $u$ and
\begin{equation}
\lambda_{s}(u)
=
\beta_{t+1}\eta_{s,t+1}
\sum_{s'=1}^S
M_{s,s'}
\frac{I_{UV,s'}(u)+I_{V,s'}(u)}{N_{s'}}.
\label{eq:cs-foi}
\end{equation}
The same contact-matrix specification, derived from a synthetic Canadian
age-contact matrix \citep{prem_projecting_2021}, is used in every region.

The parameter components evolve according to
\begin{equation}
\theta_{j,t+1}
=
\min\left\{b_j,\max\left\{a_j,\,
\theta_{j,t}+\sigma_{j,t}\varepsilon_{j,t+1}\right\}\right\},
\qquad
\varepsilon_{j,t+1}
\overset{\mathrm{iid}}{\sim}\mathcal N(0,1),
\qquad
j=1,\ldots,J,
\label{eq:cs-parameter-walk}
\end{equation}
where $\sigma_{j,t}\ge0$ is prespecified and may be zero for a component held
fixed during a transition. A step that would leave $[a_j,b_j]$ ends at the
nearest bound. We set the vaccination-rate innovation to zero
before $t_{\mathrm{vac}}$. Let
$K_{\theta,t}(d\theta_{t+1}\mid\theta_t)$ denote the kernel induced by
Eq.~\eqref{eq:cs-parameter-walk}.

Let $\Phi_t(z_t;\theta_{t+1})$ denote the one-round map given by a single
fourth-order Runge--Kutta step of Eqs.~\eqref{eq:cs-ode}--\eqref{eq:cs-foi}
over $(u_t,u_{t+1}]$, returning the five free coordinates at $u_{t+1}$. The full state transition is
\begin{equation}
\mathsf K_t
\left(
dz_{t+1},d\theta_{t+1}\mid z_t,\theta_t
\right)
=
K_{\theta,t}
\left(
d\theta_{t+1}\mid\theta_t
\right)
\delta_{\Phi_t(z_t;\theta_{t+1})}
\left(
dz_{t+1}
\right),
\label{eq:cs-transition-kernel}
\end{equation}
where $\delta_a$ is the Dirac measure at $a$. The compartments are
deterministic conditional on $\theta_{t+1}$, so the joint transition need not
have a density on the full state space. The sampling design does not enter
$\mathsf K_t$.

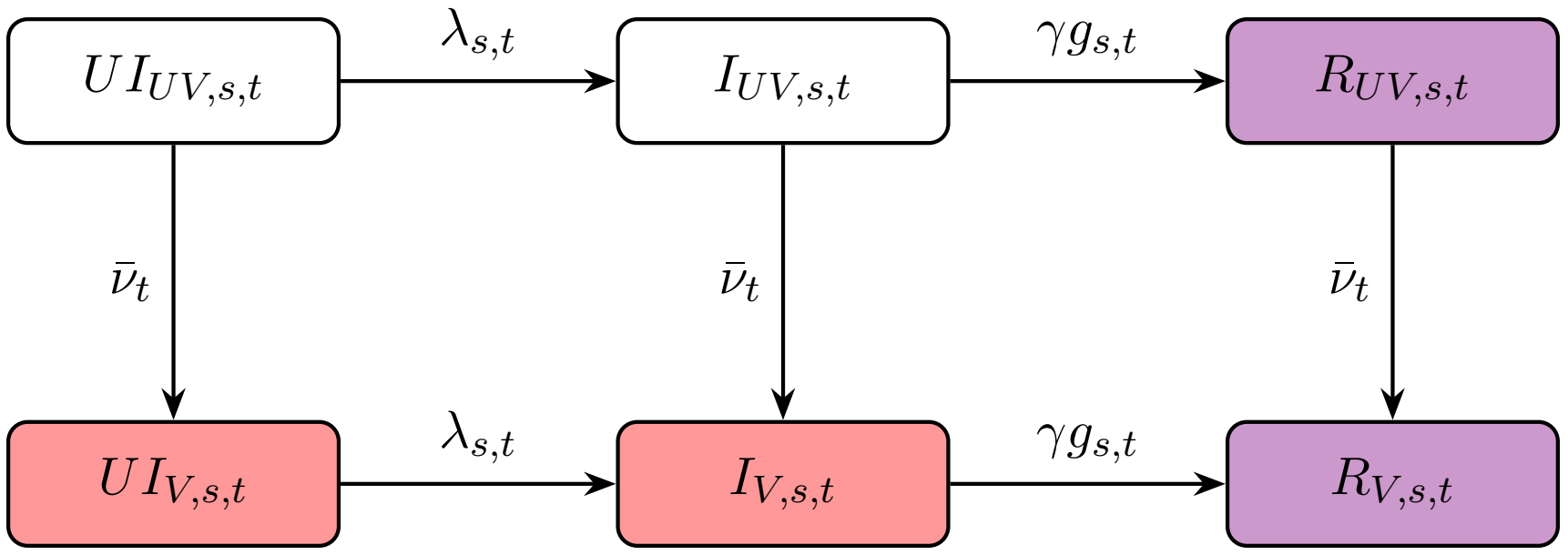
\begin{figure}[H]
\centering
\begin{tikzpicture}[
  node distance=2.0cm,
  >=Stealth,
  thick,
  every node/.style={font=\small},
  comp/.style={
    draw,
    rounded corners,
    minimum width=2.4cm,
    minimum height=0.9cm,
    align=center
  }
]
  \node[comp] (uiuv) {$UI_{UV,s,t}$};
  \node[comp, right=of uiuv] (iuv) {$I_{UV,s,t}$};
  \node[comp, fill=violet!40, right=of iuv] (ruv) {$R_{UV,s,t}$};
  \node[comp, fill=red!40, below=of uiuv] (uiv) {$UI_{V,s,t}$};
  \node[comp, fill=red!40, right=of uiv] (iv) {$I_{V,s,t}$};
  \node[comp, fill=violet!40, right=of iv] (rv) {$R_{V,s,t}$};

  \draw[->] (uiuv) -- node[above] {$\lambda_{s,t}$} (iuv);
  \draw[->] (iuv) -- node[above] {$\gamma g_{s,t}$} (ruv);
  \draw[->] (uiv) -- node[above] {$\lambda_{s,t}$} (iv);
  \draw[->] (iv) -- node[above] {$\gamma g_{s,t}$} (rv);

  \draw[->] (uiuv) -- node[left] {$\bar\nu_t$} (uiv);
  \draw[->] (iuv) -- node[left] {$\bar\nu_t$} (iv);
  \draw[->] (ruv) -- node[left] {$\bar\nu_t$} (rv);
\end{tikzpicture}
\caption{Six-compartment transmission model for each age--sex stratum.
Anti-N seropositivity is represented by
\textcolor{violet}{$R_{UV,s,t}$ and $R_{V,s,t}$}.
Anti-S seropositivity is represented by
\textcolor{violet}{$R_{UV,s,t}$ and $R_{V,s,t}$} together with
\textcolor{red}{$UI_{V,s,t}$ and $I_{V,s,t}$}.}
\label{fig:model_diagram}
\end{figure}

The rate $\gamma g_{s,t}$ simultaneously ends infectiousness and moves an
individual into a seropositive compartment. The model does not distinguish
these processes and does not include waning. The stratum-specific model-implied
anti-N seropositivity and regional estimand are
\begin{align}
\pi_{s,t,N}
&=
\frac{R_{UV,s,t}+R_{V,s,t}}{N_s},
\\
\psi_t=\varphi(x_t)
&=
\sum_{s=1}^S\omega_s\pi_{s,t,N},
\qquad
\omega_s=\frac{N_s}{\sum_{r=1}^S N_r}.
\label{eq:cs-estimand}
\end{align}
Under the non-waning model, $\psi_t$ represents cumulative seroconversion
rather than cross-sectional anti-N seroprevalence. The target width
$w^*$ applies to this model-defined estimand.

\subsection{Observation model and particle filter}
\label{sec:cs-filter}

The observation model is a joint model for the two antibody results carried by a single specimen. The compartment model implies the true joint cell probabilities
\begin{equation}
\begin{aligned}
p_{s,t}^{00}&=\frac{UI_{UV,s,t}+I_{UV,s,t}}{N_s},
&p_{s,t}^{01}&=\frac{UI_{V,s,t}+I_{V,s,t}}{N_s},\\
p_{s,t}^{10}&=0,
&p_{s,t}^{11}&=\frac{R_{UV,s,t}+R_{V,s,t}}{N_s},
\end{aligned}
\label{eq:cs-true-cells}
\end{equation}
where the first superscript is 1 if anti-N positive and the second is 1 if anti-S positive. The $10$ cell is structurally empty because recovery produces both antibodies in the latent model.

Let $\mathrm{Se}_N,\mathrm{Sp}_N,\mathrm{Se}_S,$ and $\mathrm{Sp}_S$ denote the prespecified sensitivities and specificities of the two assays. Assuming that assay errors are conditionally independent given true antibody status, the observed cell probabilities are
\begin{equation}
\begin{aligned}
q_{s,t}^{00}
&=p_{s,t}^{00}\mathrm{Sp}_N\mathrm{Sp}_S
 +p_{s,t}^{01}\mathrm{Sp}_N(1-\mathrm{Se}_S)
 +p_{s,t}^{11}(1-\mathrm{Se}_N)(1-\mathrm{Se}_S),\\
q_{s,t}^{01}
&=p_{s,t}^{00}\mathrm{Sp}_N(1-\mathrm{Sp}_S)
 +p_{s,t}^{01}\mathrm{Sp}_N\mathrm{Se}_S
 +p_{s,t}^{11}(1-\mathrm{Se}_N)\mathrm{Se}_S,\\
q_{s,t}^{10}
&=p_{s,t}^{00}(1-\mathrm{Sp}_N)\mathrm{Sp}_S
 +p_{s,t}^{01}(1-\mathrm{Sp}_N)(1-\mathrm{Se}_S)
 +p_{s,t}^{11}\mathrm{Se}_N(1-\mathrm{Se}_S),\\
q_{s,t}^{11}
&=p_{s,t}^{00}(1-\mathrm{Sp}_N)(1-\mathrm{Sp}_S)
 +p_{s,t}^{01}(1-\mathrm{Sp}_N)\mathrm{Se}_S
 +p_{s,t}^{11}\mathrm{Se}_N\mathrm{Se}_S.
\end{aligned}
\label{eq:cs-observed-cells}
\end{equation}
The assay characteristics are fixed inputs reported in Appendix Table~\ref{tab:pf_config}.

In the dual-assay phase every specimen carries both results, so $n_{s,t}=\sum_{a,b}Y_{s,t}^{ab}$ counts specimens and the paired observation model is
\begin{equation}
\mathbf Y_{s,t}\mid x_t,n_{s,t}
\sim
\operatorname{Multinomial}
\left(n_{s,t};q_{s,t}^{00},q_{s,t}^{01},q_{s,t}^{10},q_{s,t}^{11}\right),
\label{eq:cs-paired-observation}
\end{equation}
independently across strata conditional on $x_t$. After particle $i$ is propagated through the state transition in a bootstrap particle filter with $N_{\mathrm{part}}$ particles, its likelihood is
\begin{equation}
L_t^{(i)}
=
\prod_{s=1}^S
\left\{
\frac{n_{s,t}!}{\prod_{a,b}Y_{s,t}^{ab}!}
\prod_{a,b}\left(q_{s,t}^{ab,(i)}\right)^{Y_{s,t}^{ab}}
\right\}.
\label{eq:joint_loglik}
\end{equation}
The primary surveillance estimand is the latent infection-derived seroprevalence, $\psi_t$ in Eq.~\eqref{eq:cs-estimand}.

During the burn-in period, specimens are assimilated through the anti-N marginal of the same model. The model-implied probability that a specimen tests anti-N positive on an assay with sensitivity $\mathrm{Se}$ and specificity $\mathrm{Sp}$ is
\begin{equation}
q_{s,t,N}(\mathrm{Se},\mathrm{Sp})
=
\mathrm{Se}\,p_{s,t}^{11}
+
(1-\mathrm{Sp})\left(p_{s,t}^{00}+p_{s,t}^{01}\right),
\label{eq:cs-antin-marginal}
\end{equation}
which equals $q_{s,t}^{10}+q_{s,t}^{11}$ when evaluated at the Roche parameters. Writing $\mathrm{Se}_N^{\mathrm A}$ and $\mathrm{Sp}_N^{\mathrm A}$ for the Abbott assay, and letting $n_{s,t,N}^{\mathrm A}$ and $n_{s,t,N}^{\mathrm R}$ be the numbers of burn-in specimens whose anti-N result came from each platform with $y_{s,t,N}^{\mathrm A}$ and $y_{s,t,N}^{\mathrm R}$ positive, the burn-in likelihood is
\begin{equation}
L_t^{(i)}
=
\prod_{s=1}^S
\prod_{c\in\{\mathrm A,\mathrm R\}}
\operatorname{Binomial}
\left(
y_{s,t,N}^{c};
n_{s,t,N}^{c},
q_{s,t,N}^{(i)}\!\left(\mathrm{Se}_N^{c},\mathrm{Sp}_N^{c}\right)
\right).
\label{eq:cs-burnin-loglik}
\end{equation}

The bootstrap-filter weights are updated using
\begin{equation}
\widetilde w_t^{(i)}
=w_{t-1}^{(i)}\left(L_t^{(i)}\right)^\zeta,
\qquad
w_t^{(i)}
=\frac{\widetilde w_t^{(i)}}{\sum_{j=1}^{N_{\mathrm{part}}}\widetilde w_t^{(j)}}.
\label{eq:weight_update}
\end{equation}
Here $\zeta\in(0,1]$ is a likelihood exponent. With $\zeta=1$ the update is the standard Bayesian one. With $\zeta<1$ each week's likelihood is raised to a fractional power, which gives a tempered, or generalized, Bayesian update that discounts the data relative to the model \citep{bissiri_general_2016}. Tempering limits weight degeneracy, since weekly counts reach several hundred specimens per stratum and the transmission model is only an approximation. We set $\zeta=0.2$, chosen by tuning during the burn-in period. All case-study filtering intervals, and the hypothetical updates in the assurance calculation, use this tempered likelihood, so they are intervals of the tempered posterior rather than of the standard posterior.

At each week, the particles are propagated through Eq.~\eqref{eq:cs-transition-kernel}, weighted by Eq.~\eqref{eq:weight_update}, and systematically resampled when the effective sample size in Eq.~\eqref{eq:bg-ess} falls below its prespecified threshold. A week with no specimens contributes a constant likelihood, so its assimilation step reduces to prediction as in Algorithm~\ref{alg:framework}.

When the incoming observation lies far from the propagated cloud, we use the adaptive iterated particle filter of \citet{zuo_adaptive_2013}. Appendix Section~\ref{app:cs-aipf} gives the trigger rule and implementation details.

\subsection{Case-study assurance calculation and allocation}
\label{sec:cs-design}

At decision time $t$, the filtering ensemble is propagated one week to give
$\{x_{t+1}^{(i)},\bar w_{t+1}^{(i)}\}_{i=1}^{N_{\mathrm{part}}}$. Particle
$i$ implies the regional population-weighted anti-N estimand
\begin{equation}
\psi_{t+1}^{(i)}
=
\sum_{s=1}^S\omega_s\pi_{s,t+1}^{(i)},
\qquad
\pi_{s,t+1}^{(i)}
=
\frac{R_{UV,s,t+1}^{(i)}+R_{V,s,t+1}^{(i)}}{N_s}.
\label{eq:pooled_predictive_target}
\end{equation}

A candidate design is an allocation vector $\mathbf n=(n_1,\ldots,n_S)$ of specimens across strata, and its cost is the number of specimens tested,
\begin{equation}
C(\mathbf n)=\sum_{s=1}^S n_s=n.
\label{eq:cs-cost}
\end{equation}
The admissible totals $\mathcal N_{t+1}$ of Eq.~\eqref{eq:cs-admissible-totals} are fixed by a prespecified per-round floor $n_{\min}$ and ceiling $n_{\max}$ together with per-stratum bounds $n_{s,\min}$ and $n_{s,\max}$, listed in Appendix Table~\ref{tab:pf_config}. 

The design is chosen in two steps. Assurance first sets the total sample size, and a greedy rule then allocates that total across strata. For the first step, each particle's stratum cell probabilities are pooled with the population weights,
\begin{equation}
\bar q_{t+1}^{ab,(i)}
=
\sum_{s=1}^S\omega_s q_{s,t+1}^{ab,(i)},
\qquad a,b\in\{0,1\},
\label{eq:cs-pooled-cells}
\end{equation}
which are the cell probabilities of a specimen drawn at random from the regional population. Because Eq.~\eqref{eq:cs-observed-cells} is linear in the true cell probabilities, the latent anti-N cell of the pooled vector is $\sum_s\omega_s p_{s,t+1}^{11,(i)}=\psi_{t+1}^{(i)}$, so the pooled observation informs the regional estimand directly. A hypothetical sample of total size $n$ is $\mathbf Y\sim\operatorname{Multinomial}(n;\bar{\mathbf q}_{t+1}^{(i)})$ for a particle $i$ drawn from the predictive ensemble. Let $W_{\tau,t+1}(\mathbf Y,n;h_t)$ be the width of the equal-tailed $100(1-\tau)\%$ interval for $\psi_{t+1}$ after the particles are reweighted by the tempered likelihood of $\mathbf Y$. The particle-approximated assurance is
\begin{equation}
\widehat A_{t+1}(n\mid h_t)
=
\Pr_{\mathbf Y\sim\widehat m_{t+1,n}}
\left\{
W_{\tau,t+1}(\mathbf Y,n; h_t)\le w^*
\right\},
\label{eq:dynamic_assurance}
\end{equation}
which is the assurance of a population-proportional sample of size $n$.

The total is found by doubling $n$ from 1 until $\widehat A_{t+1}(n\mid h_t)\ge q$ or $n$ reaches $n_{\max}$, and then bisecting between the last two totals. Every evaluation within a round uses the same random-number seed, so totals are compared with common random numbers. Bisection assumes that assurance is nondecreasing in $n$. The search therefore returns a total $n_{t+1}^{A}$ at which estimated assurance crosses $q$, which need not be the sustained minimum of Eq.~\eqref{eq:fw-sustained-assurance}. If no total up to $n_{\max}$ reaches $q$, the round is recorded as infeasible and $n_{t+1}^{A}=n_{\max}$. The recommended total and allocation are
\begin{align}
n_{t+1}^{\mathrm{tot}}
&=
\min\left\{\max\left(n_{t+1}^{A},\,n_{\min},\,\textstyle\sum_{s}n_{s,\min}\right),\,n_{\max}\right\},
\label{eq:dynamic_assurance_sample_size}\\
\mathbf n_{t+1}^*
&=g_{t+1}(n_{t+1}^{\mathrm{tot}}; h_t).
\label{eq:cs-allocation-map}
\end{align}
The greedy rule $g_{t+1}$ starts from the stratum floors $n_{s,\min}$ and gives each further specimen to the stratum whose Beta--binomial expected interval width for $\pi_{s,t+1}$ falls most, up to the stratum ceiling $n_{s,\max}$. This proxy targets stratum-level precision and does not weight strata by population, so the allocated design generally differs from the population-proportional sample assumed in Eq.~\eqref{eq:dynamic_assurance}. The assurance value is therefore an approximation for the allocated design. Section~\ref{sec:results} reports the precision the allocated designs achieved in the replay, and Appendix Section~\ref{app:cs-assurance-computation} gives the allocation proxy and the assurance estimator.
\subsection{Retrospective replay evaluation}
\label{sec:cs-eval}

We evaluated the framework by replaying the surveillance series as a
counterfactual. At each decision time $t$, the policy used the counterfactual
history $ h_t^{\mathrm{cf}}$, containing the counterfactual
observations and the realized designs through round $t$, to select the
recommended total $n_{t+1}^{\mathrm{tot}}$ of
Eq.~\eqref{eq:dynamic_assurance_sample_size} and the allocation
$\mathbf n_{t+1}^{*}
=g_{t+1}(n_{t+1}^{\mathrm{tot}}; h_t^{\mathrm{cf}})$.

The specimens tested
in stratum $s$ during round $t+1$ formed the finite pool from which the
counterfactual sample was taken. Let $c_{s,t+1}$ denote the size of that pool
and $\mathbf y_{s,t+1}^{\mathrm{obs}}$ its realized four-cell composition.
Historical availability constrained what could be drawn, so the drawn size was
\begin{equation}
n_{s,t+1}^{\mathrm{drawn}}
=
\min\left(n_{s,t+1}^{*},c_{s,t+1}\right),
\label{eq:cs-replay-drawn}
\end{equation}
and the selected specimens were taken by simple random sampling without
replacement from the pool,
\begin{equation}
\mathbf Y_{s,t+1}^{\mathrm{cf}}
\mid
\left(\mathbf y_{s,t+1}^{\mathrm{obs}},n_{s,t+1}^{\mathrm{drawn}}\right)
\sim
\operatorname{MvHypergeom}
\left(
\mathbf y_{s,t+1}^{\mathrm{obs}},
n_{s,t+1}^{\mathrm{drawn}}
\right),
\label{eq:cs-retrospective-hypergeom}
\end{equation}
where the arguments are the pool composition and the number drawn, and the particle $i$ has round-$t+1$
likelihood
\begin{equation}
\ell_{t+1}^{(i)}
=
\prod_{s=1}^S
\operatorname{Multinomial}
\left(
\mathbf Y_{s,t+1}^{\mathrm{cf}};
n_{s,t+1}^{\mathrm{drawn}},
\mathbf q_{s,t+1}^{(i)}
\right),
\label{eq:cs-replay-likelihood}
\end{equation}
which is Eq.~\eqref{eq:joint_loglik} evaluated at the counterfactual counts
and drawn sizes. Appendix Section~\ref{app:replay-protocol} contains details on the full protocol.

\section{Case study results}
\label{sec:results}

\subsection{Analytic sample}

Table~\ref{tab:regional_data_overview} summarizes the final analytic sample and the observed anti-N and anti-S assay counts in each region. Anti-N counts cover the burn-in and the decision window. Anti-S counts cover the decision window only, since the burn-in uses anti-N alone.

\begin{table}[H]
\centering
\caption{Final analytic sample and observed assay counts by modeled region.}
\label{tab:regional_data_overview}
\begin{tabular}{lrrrrr}
\toprule
Region & Records & Unique donors & Provinces & Anti-N $n$ & Anti-S $n$ \\
\midrule
Atlantic & 98,012 & 41,665 & 4 & 98,012 & 78,515 \\
Prairies & 296,608 & 124,204 & 3 & 296,608 & 256,023 \\
BC & 170,533 & 76,108 & 1 & 170,533 & 148,659 \\
Ontario & 469,828 & 204,180 & 1 & 469,828 & 387,861 \\
\midrule
Total & 1,034,981 & 446,157 & 9 & 1,034,981 & 871,058 \\
\bottomrule
\end{tabular}
\end{table}

\subsection{Model calibration}

\begin{figure}[H]
    \centering
    \includegraphics[width=0.95\linewidth]{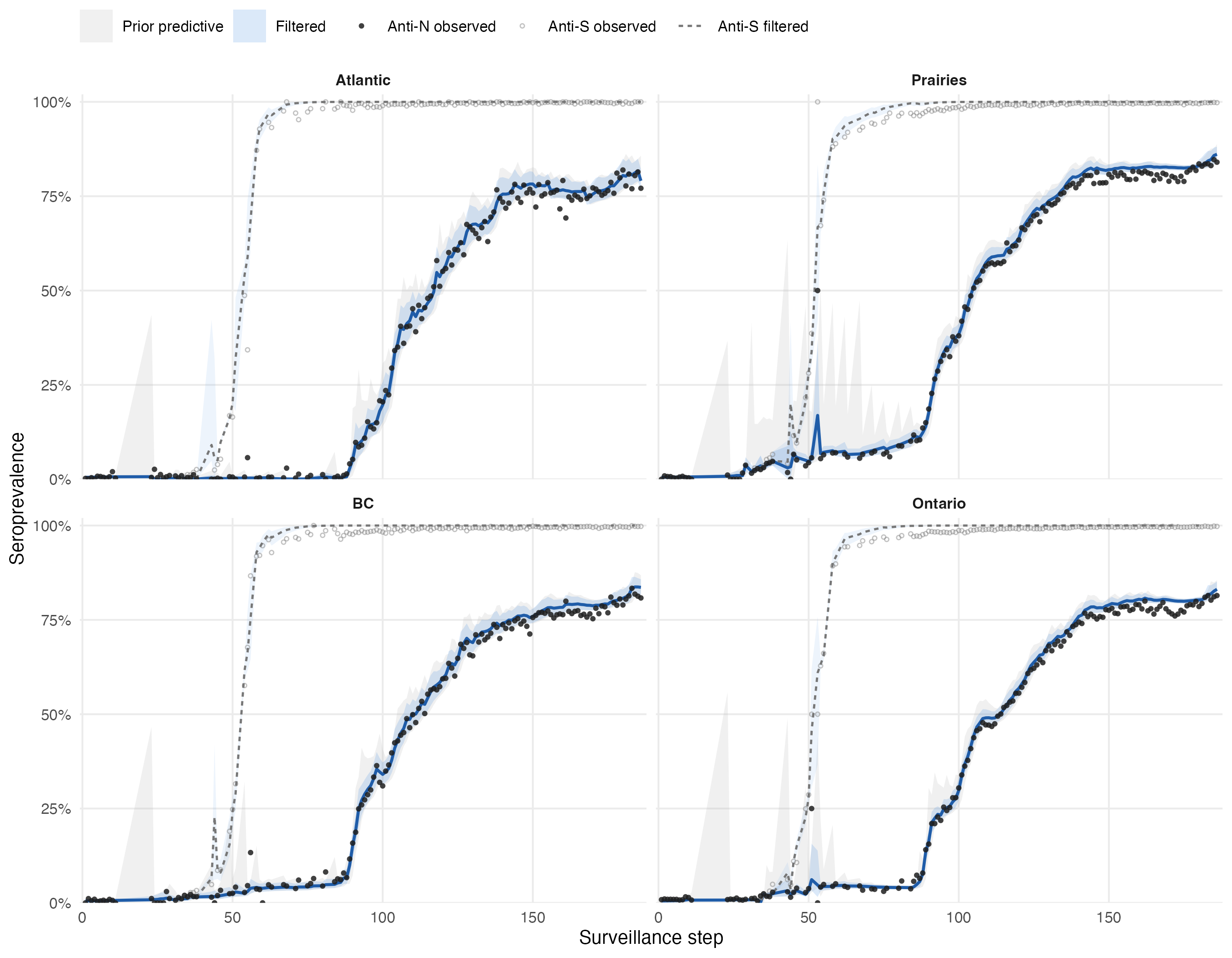}
    \caption{Pooled full-data seroprevalence trajectories by modeled region at observed anti-N surveillance steps. Gray ribbons show the anti-N one-step-ahead prior predictive interval, blue ribbons and lines show the filtered anti-N distribution after assimilating observations, and points show observed regional anti-N seroprevalence. Pale blue ribbons, dashed gray lines, and open gray points show the corresponding filtered and observed anti-S seroprevalence.}
    \label{fig:pooled_model_prior_filtered}
\end{figure}

The full-data calibration fitted each regional model with a single stratum ($S=1$) to all observed specimens. This pooled analysis reproduced the observed anti-N trajectories with RMSE values between 0.017 and 0.031 across regions, while anti-S RMSE ranged from 0.017 to 0.038 and exceeded anti-N RMSE in three of the four regions. Mean 90\% anti-N filtering interval widths were between 0.043 and 0.056,
below the design target of $w^*=0.10$.
Filter diagnostics were stable at the regional level. Mean filter ESS ranged from 635,601 to 714,779 particles, and the inferred mean $R_0$, summarized as $\beta_t/\gamma$, ranged from 1.273 to 1.391. Table~\ref{tab:pooled_model_performance} provides the region-specific summary, Figure~\ref{fig:pooled_model_prior_filtered} shows the prior predictive intervals and filtered anti-N trajectories, and the corresponding ESS and $R_0$ trajectories are shown in Appendix Figures~\ref{fig:appendix_pooled_ess} and \ref{fig:appendix_pooled_r0}.

\begin{table}[H]
\centering
\caption{Pooled full-data model calibration and filtering performance by modeled region.}
\label{tab:pooled_model_performance}
\small
\begin{tabular}{lrrrrr}
\toprule
Region & Anti-N RMSE & Anti-S RMSE & Mean width & Mean ESS & Mean $R_0$ \\
\midrule
Atlantic & 0.018 & 0.038 & 0.049 & 714,779 & 1.294 \\
Prairies & 0.031 & 0.036 & 0.056 & 668,063 & 1.273 \\
BC & 0.017 & 0.026 & 0.048 & 695,168 & 1.328 \\
Ontario & 0.021 & 0.017 & 0.043 & 635,601 & 1.391 \\
\bottomrule
\end{tabular}
\end{table}

\subsection{Counterfactual adaptive sampling performance}

\begin{figure}[H]
    \centering
    \includegraphics[width=0.95\linewidth]{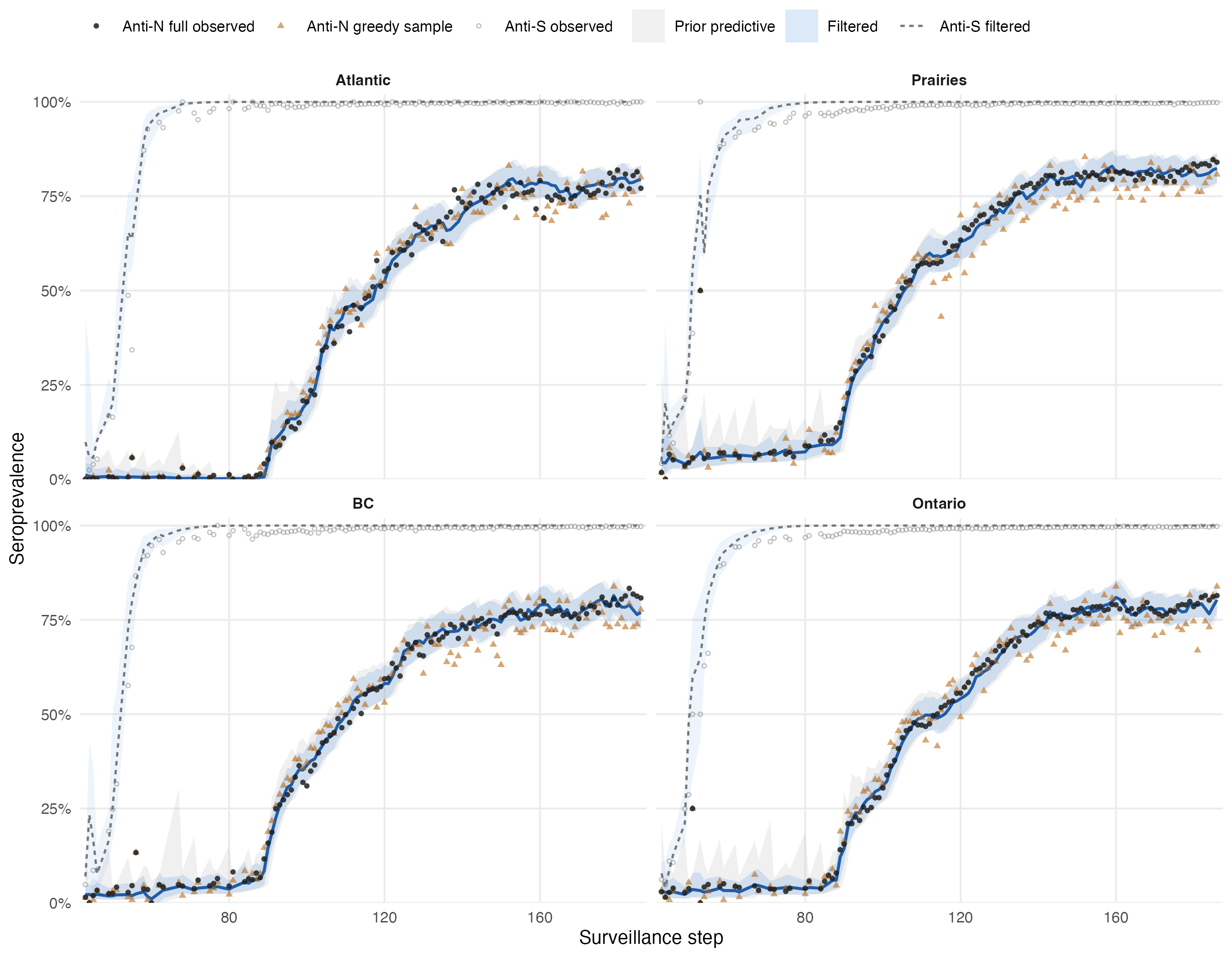}
    \caption{Counterfactual seroprevalence trajectories under the greedy adaptive sampling policy. Gray ribbons show the anti-N one-step-ahead prior predictive interval at steps with counterfactual anti-N observations, blue ribbons and lines show the filtered anti-N distribution after assimilating the greedy sample, black points show the full observed regional anti-N seroprevalence, and orange triangles show the counterfactual greedy sample seroprevalence used to update the filter. Pale blue ribbons, dashed gray lines, and open gray points show the corresponding filtered and observed anti-S seroprevalence.}
    \label{fig:greedy_counterfactual_prior_filtered}
\end{figure}

\begin{figure}[H]
    \centering
    \includegraphics[width=0.98\linewidth]{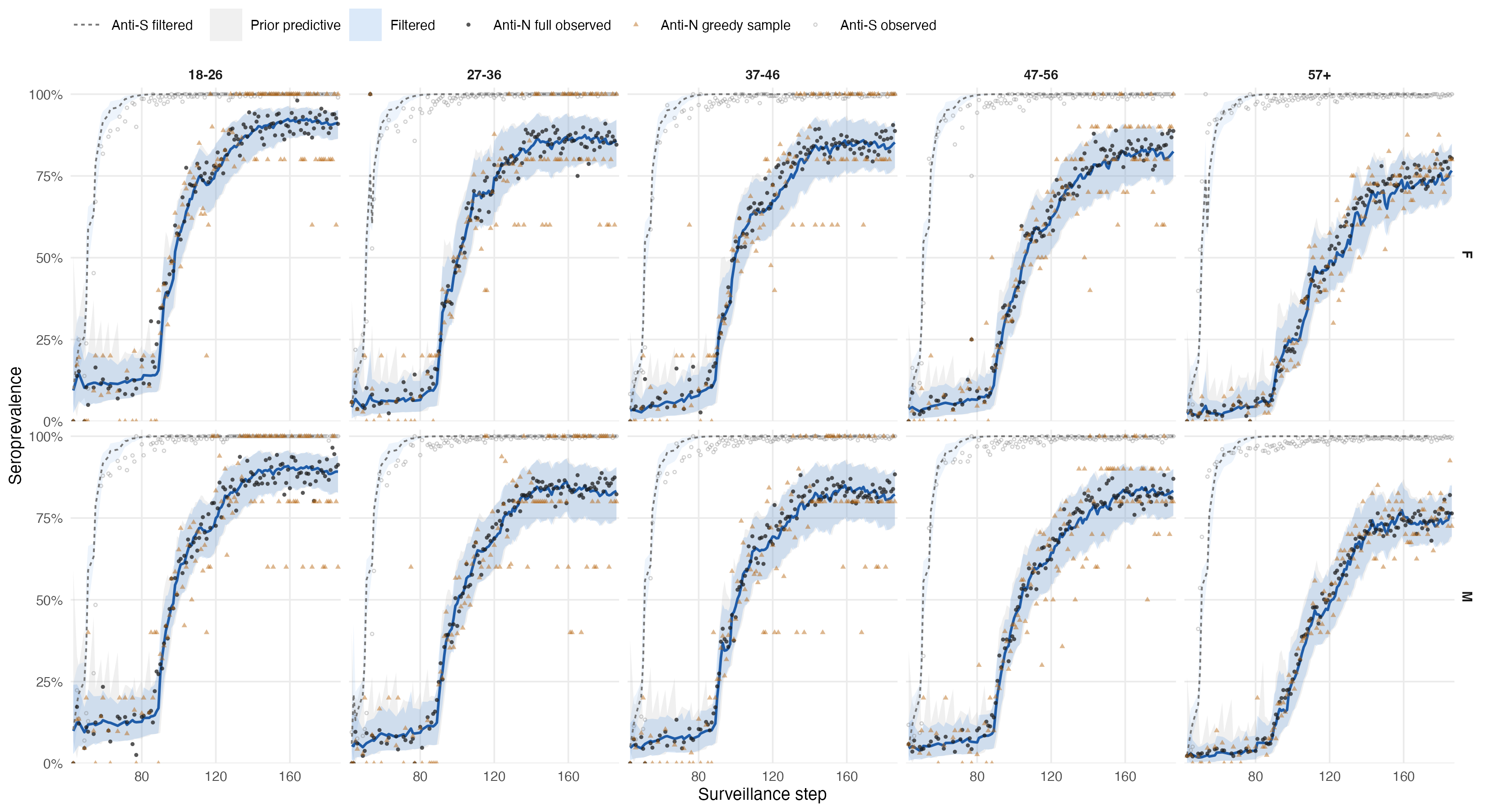}
    \caption{Prairies stratum-level anti-N trajectories in the replay, seed 1. Columns are age groups and rows are sex. Legend as in Figure~\ref{fig:greedy_counterfactual_prior_filtered}.}
    \label{fig:stratified_counterfactual_prairies}
\end{figure}

We next evaluated the counterfactual policy by replaying the surveillance series. The replay simulated no change to sampling before the week of March 1, 2021. From then on, we generated a weekly reduced subsample of specimens based on the sampling design selected by our adaptive policy, then updated the particle filter using the resulting counterfactual serological observation. Performance was then evaluated against the full observed regional seroprevalence trajectory. Across this replay window, the adaptive policy recommended 138,427 specimens and drew 128,156 after truncation to those available, subsampling from the 871,591 samples actually tested.
The program sized its testing for purposes other than this precision target, so the difference shows how far the target falls below historical volume rather than a saving over a non-adaptive version of our design. The observed total includes 533 decision-window specimens with a Roche anti-N result but no anti-S result, which Table~\ref{tab:regional_data_overview} excludes. Recommended totals tracked the one-step predictive width, with Spearman correlations of 0.85 to 0.86 across regions, and were largest between February and June 2022, during the Omicron wave. The 90\% filtering interval met the width target of 0.10 in 94\% of the 502 region-rounds with specimens, ranging from 88\% in Atlantic to 97\% in Ontario. Anti-N RMSE was between 0.020 and 0.043 across regions, with mean filtering widths between 0.075 and 0.081 (Table~\ref{tab:greedy_counterfactual_performance}; Figure~\ref{fig:greedy_counterfactual_prior_filtered}).
 The adaptive design preserved age- and sex-specific seroprevalence trajectories in the stratified counterfactual replay. Filtering performance in the stratified replay for the Prairies region is summarized in Appendix Table~\ref{tab:stratified_counterfactual_prairies} and Appendix Figure~\ref{fig:stratified_counterfactual_prairies}. The sampling effort recommended by the filter over the observation period for each of the strata is depicted in Appendix Figure~\ref{fig:stratified_counterfactual_sample_allocation}.

Table~\ref{tab:greedy_counterfactual_floor} shows how often the recommendation sat at the sampling floor or exceeded the specimens available. Over the 502 region-rounds with specimens, Eq.~\eqref{eq:dynamic_assurance_sample_size} returned the smallest admissible total of 130 specimens, the sum of the stratum floors, in 295 rounds (58.8\%), and returned a strictly larger total in the remaining 207. Rounds at the floor account for 36,690 of the 128,156 anti-N tests drawn, so 29\% of the adaptive effort reflects the operational floor rather than a binding precision requirement. The recommended total exceeded the whole regional pool in 36 rounds (7.1\%). Shortfalls were more common at the stratum level, where at least one stratum's allocation exceeded the specimens available in that stratum in 94 rounds (18.7\%), most often in Atlantic and BC. In those rounds the drawn allocation departs from the recommended one.

\begin{table}[H]
\centering
\caption{Rounds at the sampling floor and rounds limited by specimen availability in the counterfactual replay, by modeled region. Total over pool counts rounds in which the recommended total exceeded the regional pool. Stratum truncated counts rounds in which at least one stratum's allocation exceeded its pool. Counts are means over three seeds.}
\label{tab:greedy_counterfactual_floor}
\small
\begin{tabular}{lrrrrr}
\toprule
Region & With specimens & At floor & Total over pool & Stratum truncated & No specimens \\
\midrule
Atlantic & 124 & 77 (61.8\%) & 24.0 & 50.0 & 20 \\
Prairies & 126 & 74 (59.0\%) & 3.0 & 6.7 & 18 \\
BC & 127 & 72 (57.0\%) & 5.7 & 31.7 & 17 \\
Ontario & 125 & 72 (57.3\%) & 3.0 & 5.7 & 19 \\
\midrule
Total & 502 & 295 (58.8\%) & 35.7 & 94.0 & 74 \\
\bottomrule
\end{tabular}
\end{table}

\begin{table}[H]
\centering
\caption{Counterfactual filtering performance by modeled region.}
\label{tab:greedy_counterfactual_performance}
\small
\begin{tabular}{lrrrrr}
\toprule
Region & Sampling fraction & Anti-N RMSE & Anti-S RMSE & Mean width & Mean ESS \\
\midrule
Atlantic & 34.9\% & 0.024 & 0.038 & 0.075 & 615,393 \\
Prairies & 13.1\% & 0.043 & 0.039 & 0.081 & 608,771 \\
BC & 22.7\% & 0.020 & 0.029 & 0.081 & 602,067 \\
Ontario & 8.6\% & 0.026 & 0.027 & 0.080 & 605,895 \\
\bottomrule
\end{tabular}
\end{table}

\begin{figure}[H]
    \centering
    \includegraphics[width=0.95\linewidth]{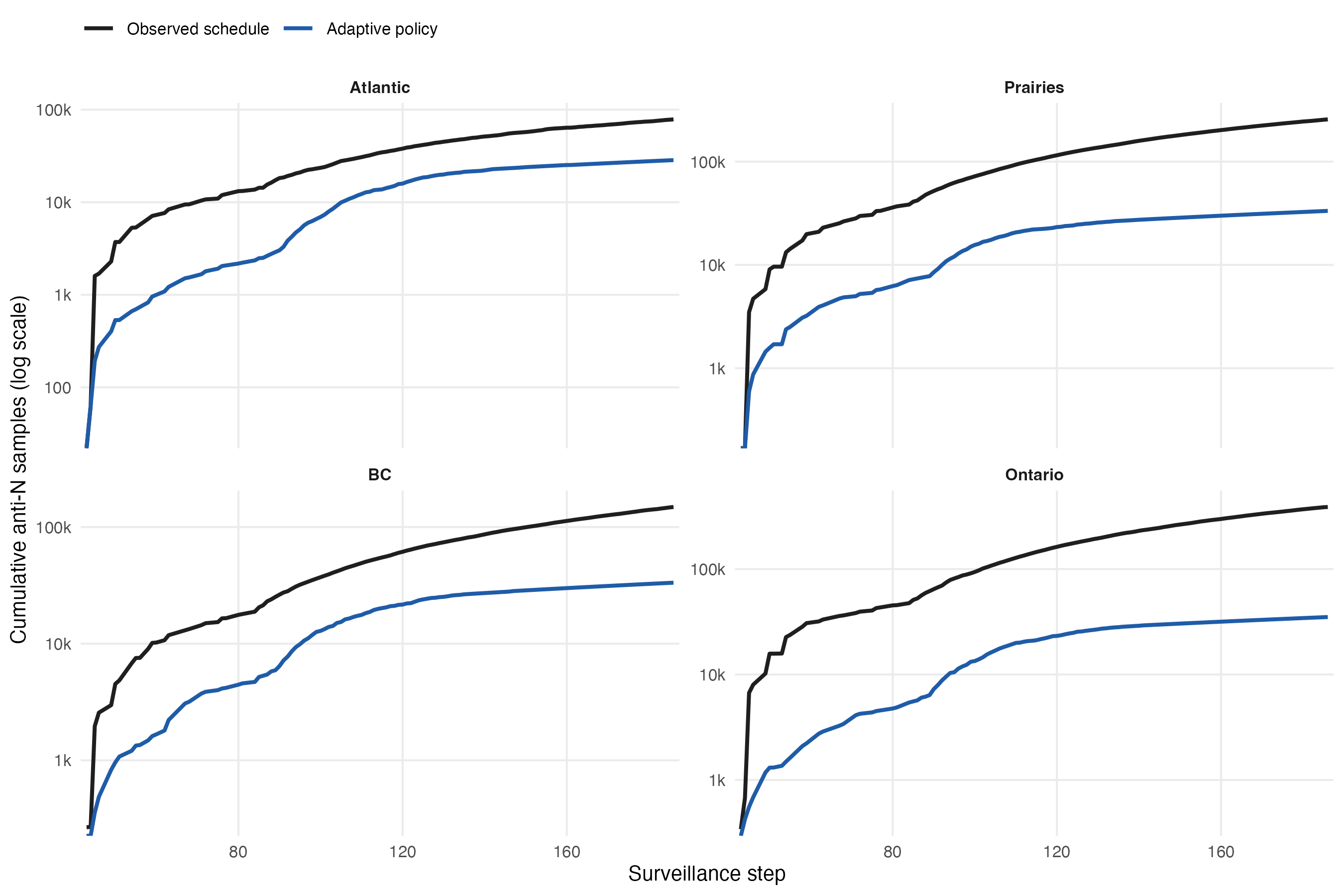}
    \caption{Cumulative anti-N sample size under the observed surveillance schedule and the adaptive policy in the regional counterfactual replay. The y-axis is log-scaled.}
    \label{fig:greedy_counterfactual_cumulative_samples}
\end{figure}

The adaptive policy drew 27,418 to 33,697 anti-N specimens per region over the replay, despite a fivefold difference in observed regional volume (Table~\ref{tab:greedy_counterfactual_efficiency}; Figure~\ref{fig:greedy_counterfactual_cumulative_samples}).

\begin{table}[H]
\centering
\caption{Cumulative anti-N sample size required by the observed and adaptive sampling schedules in the regional counterfactual replay.}
\label{tab:greedy_counterfactual_efficiency}
\small
\begin{tabular}{lrrr}
\toprule
Region & Observed anti-N $n$ & Adaptive anti-N $n$ & Sampling fraction \\
\midrule
Atlantic & 78,578 & 27,418 & 34.9\% \\
Prairies & 256,326 & 33,628 & 13.1\% \\
BC & 148,672 & 33,697 & 22.7\% \\
Ontario & 388,015 & 33,413 & 8.6\% \\
\midrule
Total & 871,591 & 128,156 & 14.7\% \\
\bottomrule
\end{tabular}
\end{table}

\section{Discussion}

We developed an adaptive sampling framework that uses a sequential predictive
distribution to select the next-round sampling design under a Bayesian assurance
criterion. When the increase in variance was anticipated, the policy raised sampling in the high-volatility rounds, and the share of intervals meeting the width target agreed with validated assurance. When the transmission rate changed without warning, realized width success fell below validated assurance in the rounds after the change, and sampling rose until the filter had relearned the transmission rate.  In the counterfactual replay, recommended sample sizes tracked the one-step
predictive uncertainty and rose during the Omicron wave, and the filtering
interval met the width target in 94\% of region-rounds. The case study
therefore shows that the design loop can run on a real surveillance series and
adapt its effort to the uncertainty carried forward from earlier rounds.

The framework differs from designs that treat each round as an independent
cross-sectional survey. The filtering distribution retains information from
earlier observations, and the process model propagates that information to the
next decision time. Additional observations have less value for posterior
precision when the resulting predictive distribution is already narrow and
more value when it is diffuse. The framework also differs from adaptive
surveillance methods that optimize sampling locations, timing, or allocation
under a fixed resource level
\citep{chipeta_adaptive_2016,case_adapting_2024,batlle_adaptive_2022,contreras_model-based_2023}.
Here the sampling effort is itself chosen through a posterior precision
requirement.

The assurance criterion gives this requirement a direct operational
interpretation. The target width $w^*$ specifies the desired posterior
precision, while $q$ specifies the predictive probability with which that
precision should be achieved. Both should reflect the intended surveillance
use and the consequences of collecting too little or too much information. The
assurance remains conditional on the fitted model and the observed history. It
is a predictive probability that the interval will meet the width target, not
a guarantee of that event and not a statement about interval coverage. The
credibility level $1-\tau$ fixes which interval is used, while $q$ fixes how
probable a narrow enough interval must be, even though both are 0.90 here.

Sustained assurance addresses a separate feature of discrete designs.
Requiring the constraint to hold at the selected size and at all larger
admissible sizes avoids relying on an isolated first crossing of the threshold,
which the shifting outcome lattice makes possible. The simulations used this
rule. The case study used a simpler approximation. It set the total by
assurance for a population-proportional sample, found by bisection, and then
allocated that total with a greedy proxy, so its assurance is approximate for
the allocated design. Evaluating assurance over the stratified allocation
family with the sustained rule would remove this approximation.

The usefulness of the recommended sampling designs depends on the adequacy of the
predictive distribution. Additional data can reduce uncertainty represented
by the model, but the assurance criterion does not account for misspecification
that is absent from that distribution. Prospective use therefore requires
predictive model checking, sensitivity analyses for the process and
observation models, and monitoring of particle and Monte Carlo error. These
safeguards are compatible with the general
framework, as are alternative process models, estimands, precision
functionals, allocation rules, and sampling costs.
The simultaneous lower-bound rule used in the simulations addresses finite
predictive Monte Carlo error conditional on the particle ensemble.

The simulation study in Section~\ref{sec:sim-unannounced} quantifies the resulting error. When the
transmission rate changed without warning, validated assurance stayed at 0.960 while
realized width success fell to 0.894, so the criterion overstated the achieved
precision in the period when the epidemic was least well understood. The
assurance value cannot detect the departure that degrades it, because it is a
predictive probability computed under the model that omits the departure.
Filter diagnostics such as the effective filter sample size, which fell markedly
in the transition window, are what signal an unreliable predictive distribution
in real time. The filter recovers, because observations that fall outside its
high-probability region widen the next predictive distribution and raise the
selected total. How quickly this self correction happens depends on the model and on how soon
a change is reflected in the observed data. In our simulation, realized width success
was close to validated assurance again once the transition window had passed (0.965
against 0.971). The delay reflects that a jump in transmission rate appears in
seroprevalence only after infection and seroconversion. A minimum sampling floor and routine predictive checking are
therefore complements to the criterion. 

The case-study findings remain subject to several application-specific
limitations. Historical specimen availability capped
each allocation, so larger counterfactual designs could not be evaluated. The
recommended total exceeded the regional pool in about 7\% of rounds, and at
least one stratum's allocation exceeded its pool in about 19\%. The crude observed proportion
used for RMSE and empirical inclusion is weighted by the donor sample rather
than by the stratum populations that define the estimand, and it contains the
counterfactual subsample, so it is not an independent validation outcome.
The blood donor population is also not fully representative of the general population, although this limitation applies similarly to other surveillance populations \citep{knight_sociodemographic_2025}. The transmission model omits reinfection and antibody
waning, while the observation model does not represent recurring donors or the
finite-population correction implied by subsampling a fixed specimen pool. Individual parameters of the case-study model were not identifiable from surveillance data alone, since the
transmission rate and the susceptibility modifiers enter the force of infection
only through their product, so the model was built to give accurate predictive
distributions rather than accurate  estimates of SARS-CoV-2 epidemiological parameters. All parameters were treated as
time-varying with fixed random-walk standard deviations. These standard
deviations, the initial susceptibility modifiers, and the age-graded stratum
floors were fixed before the reported runs but were chosen during model
development from trial runs on the same historical series, so the replay is not
an out-of-sample test of these settings. A potential improvement would be to estimate
these volatilities from the data, using methods such as particle Markov
chain Monte Carlo \citep{dureau_capturing_2013}.

Prospective evaluation is needed to determine whether the adaptive behavior and the precision it achieves persist when designs are implemented before specimens are
collected. If it does, surveillance programs could maintain regional
monitoring with less testing or redirect samples toward subgroup estimates,
additional assays, or complementary data streams. The same design principle
could be applied to other pathogens when a suitable predictive model can be
updated as surveillance observations arrive.

\section*{Acknowledgments}

We thank Anna Heath (The Hospital for Sick Children and the Dalla Lana School
of Public Health, University of Toronto) for helpful suggestions on Bayesian
assurance.

\section*{Funding}

This research was funded in part by Canadian Blood Services, the Natural Sciences and Engineering
Research Council of Canada (NSERC) [funding reference number RGPIN-2023-04160,
PI W. Alton Russell] and by a Steinburg Seed Grant from the McGill Global Health Programs and the McGill Interdisciplinary Initiative in Infection and Immunity (MI4) [PI W. Alton Russell]

\section*{Author contributions}

\textbf{SM.} Conceptualization, methodology, software, formal analysis, writing
(original draft, review and editing).
\textbf{NO.} Conceptualization, methodology, writing (review and editing).
\textbf{SFO.} Conceptualization, methodology, resources, writing (review
and editing).
\textbf{DLB.} Conceptualization, methodology, resources, writing
(review and editing).
\textbf{WAR.} Conceptualization, methodology, funding acquisition, supervision,
writing (review and editing).

\newpage

\section*{References}
\printbibliography[heading=none]

\newpage

\section*{Appendix}

\subsection*{Additional tables and figures}
\phantomsection
\label{app:additional-results}

\begin{figure}[H]
    \centering
    \includegraphics[width=0.9\linewidth]{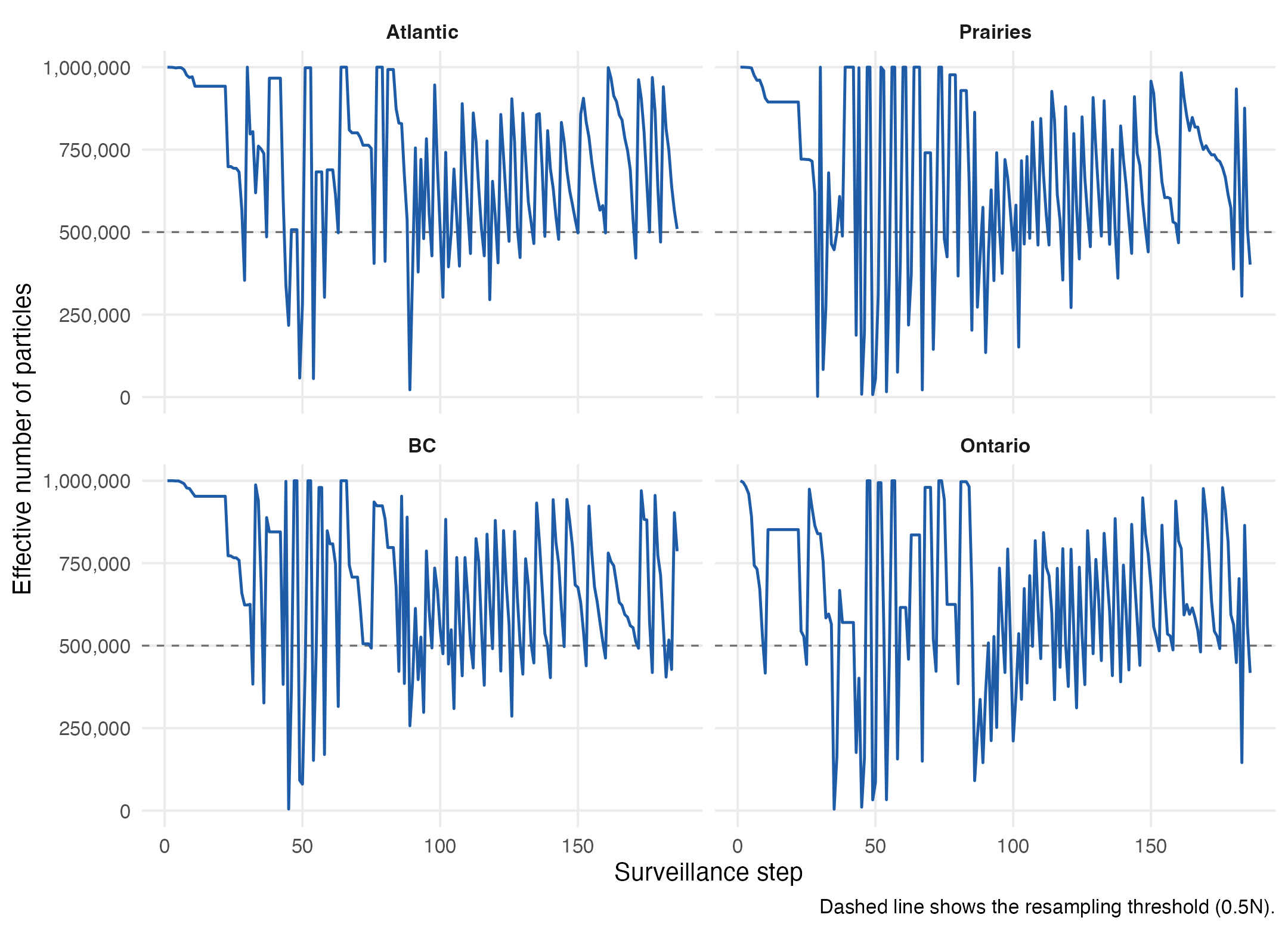}
    \caption{Effective filter sample size over time in the pooled full-data model calibration. The dashed line marks the resampling threshold.}
    \label{fig:appendix_pooled_ess}
\end{figure}

\begin{figure}[H]
    \centering
    \includegraphics[width=0.9\linewidth]{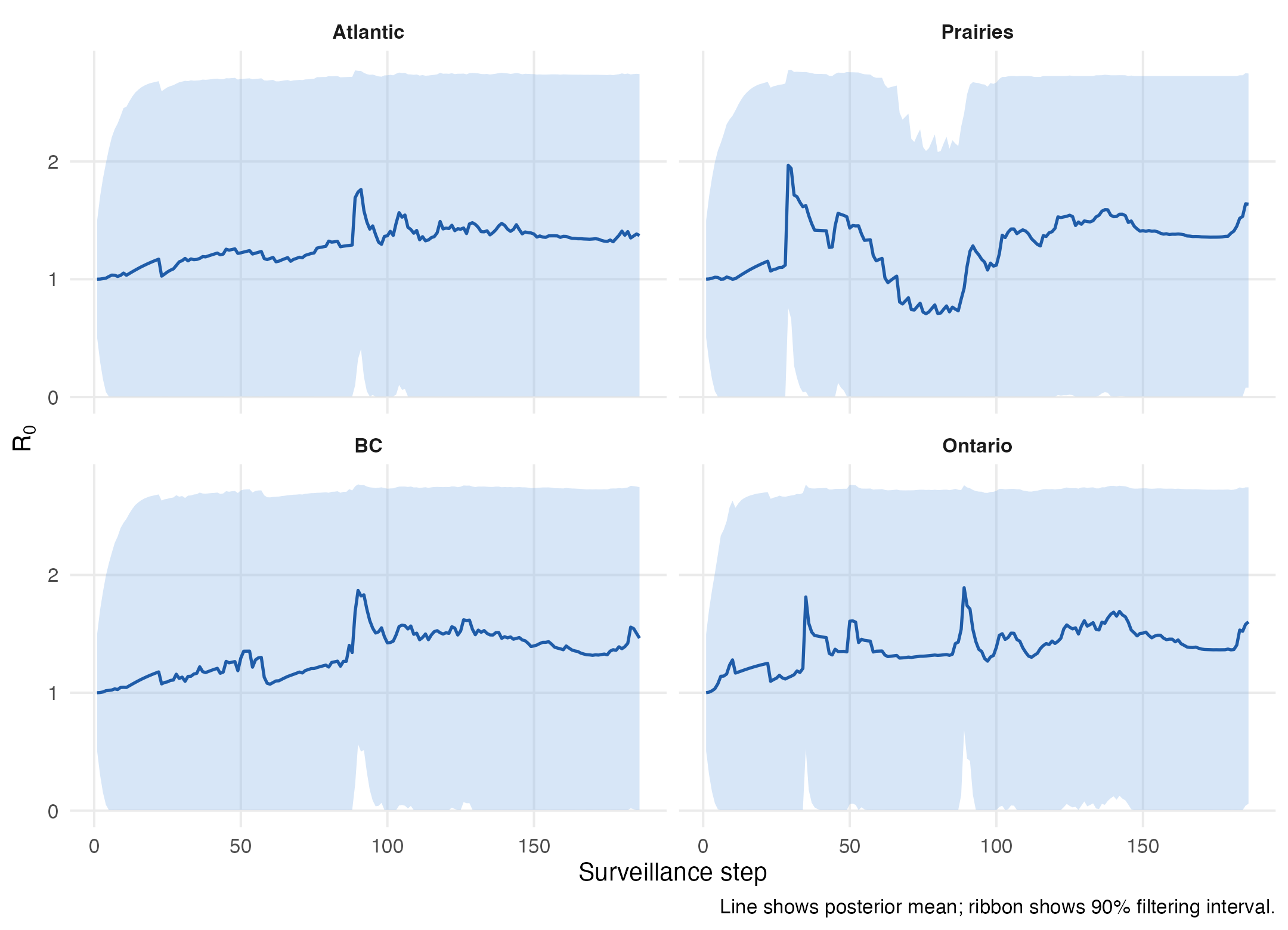}
    \caption{Posterior mean $R_0$ over time in the pooled full-data model calibration. Ribbons show 90\% filtering intervals.}
    \label{fig:appendix_pooled_r0}
\end{figure}

\begin{table}[H]
\centering
\caption{Case-study filter and design settings.}
\label{tab:pf_config}
\small
\begin{tabular}{lll}
\toprule
\textbf{Parameter} & \textbf{Value} & \textbf{Description} \\
\midrule
\multicolumn{3}{l}{\textit{Filter}} \\
$N$ & 1,000,000 & Number of particles \\
$s_T$ & $0.5N$ & Resampling threshold on the effective number of particles \\
Parameter process & Random walk & Gaussian random walk on the natural scale, clipped to bounds \\
Noise scale & 200 & Multiplier on the random-walk SDs below \\
$\sigma_\beta$ & 0.0003 / 0.0005 & Transmission-rate SD (calibration / replay) \\
$\sigma_\nu$ & 0.025 & Vaccination-rate SD \\
$\sigma_\eta$ & 0.002 & Susceptibility-modifier SD \\
$\sigma_g$ & 0.0025 & Recovery-modifier SD \\
$[a_\beta,b_\beta]$ & $[0.001,\,0.55]$ & Transmission-rate bounds \\
$[a_\nu,b_\nu]$ & $[0.05,\,0.55]$ & Vaccination-rate bounds \\
$[a_\eta,b_\eta]$ & $[0,\,100]$ & Susceptibility-modifier bounds \\
$[a_g,b_g]$ & $[0.01,\,100]$ & Recovery-modifier bounds \\
$\gamma$ & Fixed & Recovery rate, no random walk \\
$\eta_{s,0}$ & 0.3 / 0.9 / 1.0 & Initial susceptibility modifier for 57+ / 47--56 / younger strata \\
$t_{\mathrm{vac}}$ & Week 43 & First vaccination round \\
$\zeta$ & 0.2 & Likelihood exponent in Eq.~\eqref{eq:weight_update}, calibration and replay \\
$\rho_{\mathrm{AIPF}}$ & 2 & AIPF trigger threshold on $\Delta_t^2$, active from week 18 \\
$K$ & 4 & AIPF maximum iterations \\
\midrule
\multicolumn{3}{l}{\textit{Observation model}} \\
$\mathrm{Se}_N^{\mathrm A},\ \mathrm{Sp}_N^{\mathrm A}$ & 0.927, 0.999 & Abbott anti-N, burn-in only \citep{national_serology_group_2020} \\
$\mathrm{Se}_N,\ \mathrm{Sp}_N$ & 0.972, 0.998 & Roche anti-N \citep{national_serology_group_2020} \\
$\mathrm{Se}_S,\ \mathrm{Sp}_S$ & 0.979, 0.9995 & Roche anti-S \citep{riester_performance_2021} \\
\midrule
\multicolumn{3}{l}{\textit{Design}} \\
$1-\tau$ & 0.90 & Credibility level of equal-tailed intervals \\
$w^*$ & 0.10 & Target interval width \\
$q$ & 0.90 & Assurance level \\
$n_{\min}$, $n_{\max}$ & 100, 10,000 & Per-round total bounds \\
$n_{s,\min}$ & 40 / 10 / 5 & Stratum floor for 57+ / 47--56 / younger strata, summing to 130 \\
$n_{s,\max}$ & 2,000 & Stratum ceiling \\
$R_{\mathrm{MC}}$ & 750 & Predictive draws per evaluated total \\
$N_{\mathrm{sup}}$ & 1,000 & Particles in the assurance support \\
Seeds & 3 & Replay repetitions per region \\
\bottomrule
\end{tabular}
\end{table}

\begin{table}[H]
\centering
\caption{Prairies stratum-level replay performance, means over three seeds. The width target applies to the regional estimand.}
\label{tab:stratified_counterfactual_prairies}
\small
\begin{tabular}{lrrrr}
\toprule
Stratum & Sampling fraction (\%) & Anti-N RMSE & Mean width & Mean particle ESS \\
\midrule
F 18--26 & 17.6 & 0.042 & 0.143 & 602,616 \\
M 18--26 & 20.9 & 0.042 & 0.147 & 602,616 \\
F 27--36 & 14.5 & 0.091 & 0.159 & 605,657 \\
M 27--36 & 10.7 & 0.034 & 0.164 & 605,779 \\
F 37--46 & 14.2 & 0.039 & 0.160 & 602,616 \\
M 37--46 & 10.2 & 0.037 & 0.166 & 602,616 \\
F 47--56 & 17.3 & 0.040 & 0.162 & 602,616 \\
M 47--56 & 11.0 & 0.032 & 0.162 & 602,616 \\
F 57+   & 16.7 & 0.036 & 0.159 & 605,657 \\
M 57+   & 9.6 & 0.026 & 0.155 & 602,616 \\
\bottomrule
\end{tabular}
\end{table}

\begin{figure}[H]
    \centering
    \includegraphics[width=0.98\linewidth]{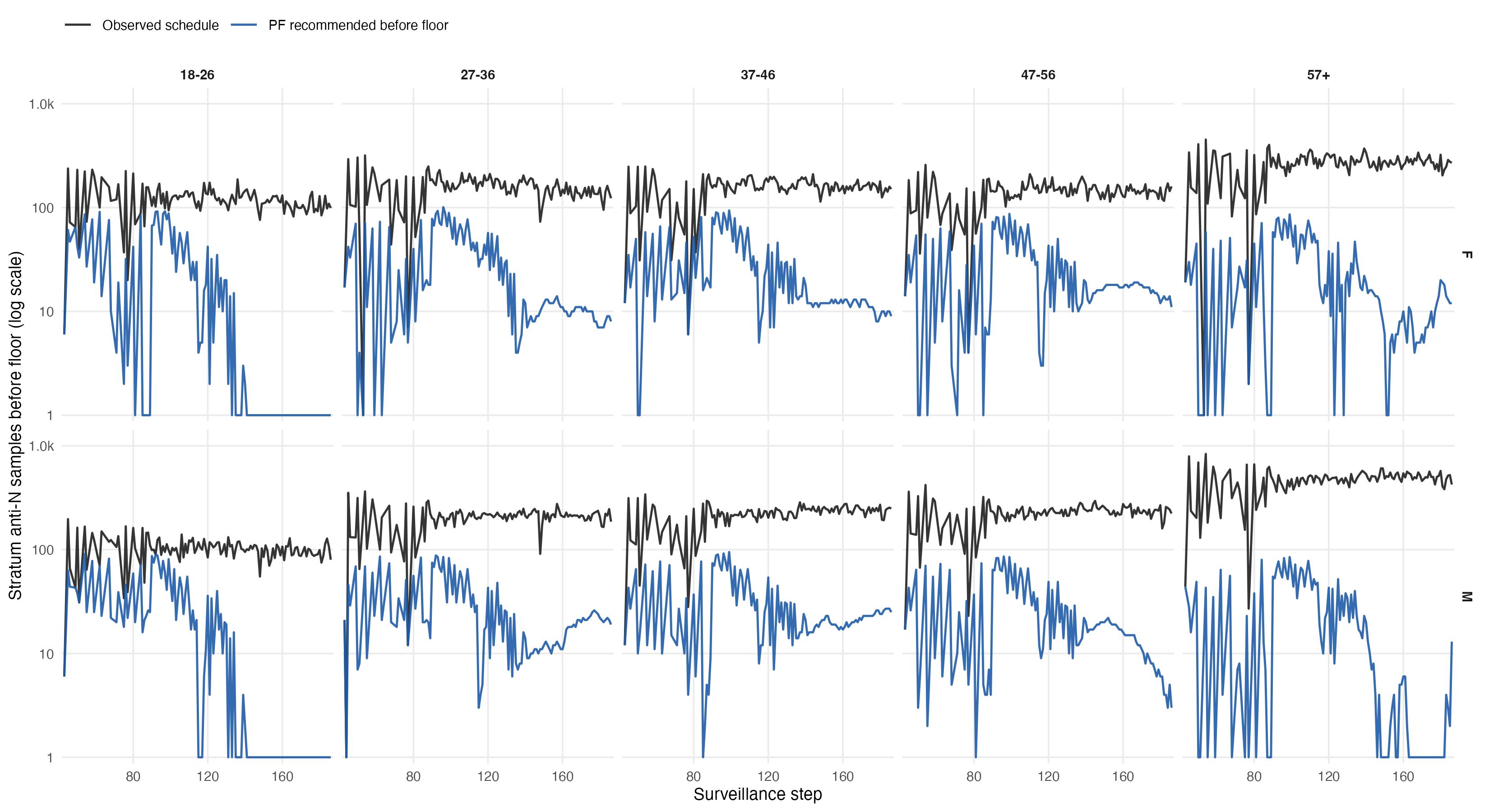}
    \caption{Observed and recommended anti-N specimens by Prairies stratum, seed 1, before the stratum floor. Log scale.}
    \label{fig:stratified_counterfactual_sample_allocation}
\end{figure}

\subsection*{Known-change simulation}
\phantomsection
\label{app:simulation-details}

\paragraph{Model and filter.}
The state in Eq.~\eqref{eq:sim-state} started from mean
$\mu_0=(\operatorname{logit}0.12,\operatorname{logit}0.28)^\top$ and covariance
$\Sigma_0=D_0RD_0$, with $D_0=\operatorname{diag}(0.30,0.30)$ and $R$ the
correlation matrix with off-diagonal 0.45. The ordinary innovation covariance
was $\Sigma=DRD$ with $D=\operatorname{diag}(0.10,0.13)$, and $a_t=3$ in rounds
6 and 7 and 1 otherwise. The same values generated the truth and propagated
the 1000 particles. Each replicate drew its own trajectory, observations and
particles. Particles were weighted by the binomial likelihood of
Eq.~\eqref{eq:sim-observation} and systematically resampled when the effective
number of particles fell below 500.

\paragraph{Allocation family.}
For stratum $s$ let
$c_{s}=\omega_s^2\sum_i\bar w_{t+1}^{(i)}p_{s,t+1}^{(i)}\{1-p_{s,t+1}^{(i)}\}$,
so that $\sum_s c_s/n_s$ approximates the sampling variance of $\psi_{t+1}$.
Starting from $(n_A,n_B)=(5,5)$, each further observation went to the stratum
with the largest reduction $c_s/\{n_s(n_s+1)\}$, with ties broken by a fixed
order. The allocations at totals $20,40,\ldots,240$ formed the candidate
family. This proxy built the family but did not enter the assurance
calculation.

\paragraph{Assurance and selection.}
Design-stage assurance used $R_{\mathrm D}=1000$ predictive draws. Each draw
sampled one particle from the predictive weights and generated the counts for
every candidate from one shared sequence of uniform variables, so candidates
were compared with common random numbers while each kept its own predictive
distribution. For each draw the particles were reweighted by the binomial
likelihood of the hypothetical counts and the width of the 90\% interval of
$\psi_{t+1}$ was recorded. With $X$ passing draws out of $R_{\mathrm D}$ and
$K=12$ candidates, each candidate received the one-sided Clopper--Pearson lower
bound $F_{\mathrm{Beta}}^{-1}(0.05/K;X,R_{\mathrm D}-X+1)$, which gives
simultaneous confidence of at least 0.95 by the Bonferroni inequality. The
selected total was the smallest whose backward cumulative minimum of these
bounds was at least 0.90. An independent set of $R_{\mathrm V}=5000$ draws
re-estimated the assurance of the selected design without altering the choice.
This rule protects the selection against predictive Monte Carlo error but not
against particle error or misspecification.

\paragraph{Performance measures.}
Measures were computed within each of the 100 replicates and then averaged.
Coverage was the fraction of rounds in which the 90\% predictive or posterior
interval contained $\psi_t$, and width success the fraction with posterior
width at most 0.10. Validation calibration was width success minus mean
validated assurance. Design response was the within-trajectory Spearman
correlation between predictive width and selected total, and allocation
agreement the mean absolute difference between the allocated share of stratum
$A$ and $\sqrt{c_A}/(\sqrt{c_A}+\sqrt{c_B})$. Monte Carlo intervals were the
mean $\pm1.96$ standard errors across replicates, truncated to the admissible
range.

\begin{table}[H]
\centering
\caption{Simulation summary across replicated trajectories. Intervals are normal-approximation 95\% Monte Carlo intervals for the replicate-level mean.}
\label{tab:simple_simulation_summary}
\footnotesize
\begin{tabular}{llrr}
\toprule
Component & Measure & Estimate & 95\% interval \\
\midrule
Filtering & Predictive interval coverage & 0.903 & (0.882, 0.925) \\
Filtering & Posterior interval coverage & 0.913 & (0.889, 0.937) \\
Filtering & Posterior RMSE & 0.025 & (0.023, 0.026) \\
Design response & Mean sample size in ordinary rounds & 45.4 & (42.5, 48.4) \\
Design response & Mean sample size in high-volatility rounds & 143.4 & (137.1, 149.7) \\
Design response & Predictive-width and effort correlation & 0.864 & (0.848, 0.881) \\
Allocation & Allocation-share mean absolute deviation & 0.007 & (0.007, 0.008) \\
Assurance & Validated assurance & 0.974 & (0.973, 0.976) \\
Assurance & Sustained simultaneous lower bound & 0.960 & (0.958, 0.962) \\
Assurance & Realized width success & 0.970 & (0.961, 0.979) \\
Assurance & Validation calibration difference & -0.004 & (-0.013, 0.005) \\
Assurance & Selected designs below target after validation & 0.000 & (0.000, 0.000) \\
Sustained assurance & Fraction of nonmonotone round curves & 0.336 & (0.306, 0.366) \\
Sustained assurance & Mean maximum raw-sustained gap & 0.001 & (0.001, 0.001) \\
Numerical & Mean ESS fraction & 0.571 & (0.559, 0.582) \\
Numerical & Design-validation assurance error & 0.004 & (0.003, 0.004) \\
Efficiency & Total effort over 12 rounds & 741.2 & (701.8, 780.6) \\
\bottomrule
\end{tabular}
\end{table}

\subsection*{Unanticipated-change simulation}
\phantomsection
\label{app:sir-variant}

\paragraph{Model and filter.}
The recovery rate was $\gamma=0.5$ per round and the contact matrix had rows
$(1.20,0.35)$ and $(0.25,0.90)$, with spectral radius $\rho(M)=1.382$, so that
$R_0=\beta\rho(M)/\gamma$. Each round advanced Eq.~\eqref{eq:sir-system} by ten
exponential-Euler substeps. A bootstrap filter with 1200 particles tracked
$(S_s,I_s,R_s)$ and $\beta_t$, initialized by perturbing
$\operatorname{logit}I_{s,0}$, $\operatorname{logit}R_{s,0}$ and $\log\beta_0$
with standard deviations 0.25, 0.15 and 0.10, and resampled systematically
when the effective number of particles fell below 600. The filter's
random-walk SD on $\log\beta$ was 0.10, against 0.04 in the data-generating
process. The larger value lets the filter relearn the jump but makes it
conservative before the jump, so the recovery in
Section~\ref{sec:sim-unannounced} is a property of this setting.

\paragraph{Design.}
Candidate totals were
$\{20,40,\ldots,200\}\cup\{240,280,\ldots,600\}\cup\{700,800,\ldots,1600\}$,
with a floor of ten per stratum. Allocation, common random numbers,
lower-bound selection and validation followed
Appendix Section~\ref{app:simulation-details}, with 600 design draws and 3000
validation draws.

\paragraph{Recovery measures.}
Transmission-rate recovery was the first round from $t_{\mathrm v}$ after which
the posterior mean of $\beta$ was within 15\% of the truth for two consecutive
rounds. Trajectory recovery was the first round after the predictive first
missed $\psi_t$, by excluding it or by a mean error above 0.010, at which the
posterior interval covered $\psi_t$ and the posterior mean was within 0.015 for
two consecutive rounds.

\paragraph{Coverage by period.}
Before the jump, predictive, posterior and transmission-rate coverage were
0.983, 0.973 and 0.999, reflecting the larger filter random walk. Across the
transition window, rounds 12 to 16, they fell to 0.836, 0.844 and 0.508, and
the relative error of the posterior mean of $\beta$ rose from 0.081 to 0.271.
The $\beta$ interval covered the truth in 0.24 of replicates in the jump round
and in more than 0.90 from round 17. After the window, coverage returned to
0.922, 0.920 and 0.992. Design-stage and validation assurance differed by 0.005
on average, and 0.481 of round-level raw assurance curves were nonmonotone.

\begin{table}[H]
\centering
\caption{Stratified SIR simulation with an unanticipated change in transmission. Summaries are across 100 replicated trajectories. Intervals are normal-approximation 95\% Monte Carlo intervals for the replicate-level mean.}
\label{tab:sir_variant_summary}
\footnotesize
\begin{tabular}{llrr}
\toprule
Component & Measure & Estimate & 95\% interval \\
\midrule
Interval coverage & Predictive coverage, Before the jump & 0.983 & (0.974, 0.991) \\
 & Predictive coverage, Transition window & 0.836 & (0.793, 0.879) \\
 & Predictive coverage, After the window & 0.922 & (0.909, 0.936) \\
 & Posterior coverage, Before the jump & 0.973 & (0.961, 0.984) \\
 & Posterior coverage, Transition window & 0.844 & (0.806, 0.882) \\
 & Posterior coverage, After the window & 0.920 & (0.905, 0.935) \\
Transmission recovery & Transmission-rate coverage, Before the jump & 0.999 & (0.997, 1.000) \\
 & Transmission-rate coverage, Transition window & 0.508 & (0.454, 0.562) \\
 & Transmission-rate coverage, After the window & 0.992 & (0.986, 0.997) \\
 & Transmission-rate relative error, Before the jump & 0.081 & (0.076, 0.086) \\
 & Transmission-rate relative error, Transition window & 0.271 & (0.260, 0.282) \\
 & Transmission-rate relative error, After the window & 0.099 & (0.092, 0.105) \\
Design response & Mean selected total, Before the jump & 88.6 & (83.7, 93.5) \\
 & Mean selected total, Transition window & 359.2 & (331.5, 386.9) \\
 & Mean selected total, After the window & 381.3 & (360.5, 402.0) \\
 & Effort relative to fixed-sample rule, Before the jump & 0.629 & (0.599, 0.659) \\
 & Effort relative to fixed-sample rule, Transition window & 0.615 & (0.596, 0.633) \\
 & Effort relative to fixed-sample rule, After the window & 0.436 & (0.419, 0.453) \\
Assurance & Realized width success, Before the jump & 0.988 & (0.982, 0.994) \\
 & Realized width success, Transition window & 0.894 & (0.869, 0.919) \\
 & Realized width success, After the window & 0.965 & (0.956, 0.974) \\
 & Validated assurance, Before the jump & 0.977 & (0.977, 0.978) \\
 & Validated assurance, Transition window & 0.960 & (0.958, 0.962) \\
 & Validated assurance, After the window & 0.971 & (0.969, 0.973) \\
Recovery timing & Transmission-rate recovery lag in rounds & 5.2 & (4.8, 5.6) \\
 & Trajectory recovery lag in rounds & 2.8 & (2.3, 3.2) \\
Numerical & Mean effective-sample-size fraction & 0.659 & (0.654, 0.665) \\
 & Design-validation assurance error & 0.005 & (0.005, 0.005) \\
 & Fraction of nonmonotone round curves & 0.481 & (0.462, 0.501) \\
 & Fraction of infeasible decisions & 0.000 & (0.000, 0.000) \\
\bottomrule
\end{tabular}
\end{table}

\subsection*{Case-study adaptive iterated particle filter}
\phantomsection
\label{app:cs-aipf}

The bootstrap update can collapse when an observation lies far from the
propagated particles. In such rounds we use the adaptive iterated particle
filter of \citet{zuo_adaptive_2013}. The trigger compares the regional observed
proportions $\widetilde{\mathbf p}_t=(\widetilde p_{t,N},\widetilde p_{t,S})^\top$
with their predictive mean $\widehat{\mathbf m}_t$ and covariance
$\widehat\Sigma_t$, which combines spread across particles with the multinomial
covariance implied by Eq.~\eqref{eq:cs-observed-cells}. The AIPF runs when
\begin{equation}
\Delta_t^2
=
(\widetilde{\mathbf p}_t-\widehat{\mathbf m}_t)^{\!\top}
\widehat\Sigma_t^{-1}
(\widetilde{\mathbf p}_t-\widehat{\mathbf m}_t)
>
\rho_{\mathrm{AIPF}},
\label{eq:aipf_trigger}
\end{equation}
with $\rho_{\mathrm{AIPF}}$ referenced to a $\chi^2_2$ distribution. It then
redraws and moves particles toward the observation under the annealing
schedule of \citet{zuo_adaptive_2013}, and the final iteration uses the
tempered four-cell likelihood of Eqs.~\eqref{eq:joint_loglik}
and~\eqref{eq:weight_update}. Settings are in Table~\ref{tab:pf_config}.

\subsection*{Case-study allocation and assurance computation}
\phantomsection
\label{app:cs-assurance-computation}

\paragraph{Allocation family.}
For stratum $s$, the predictive mean and variance of $\pi_{s,t+1}$ over the
particles, clipped away from the Bernoulli bounds, define a working
$\operatorname{Beta}(a_s,b_s)$ with $a_s+b_s=\mu_s(1-\mu_s)/v_s-1$. The
allocation proxy $\overline W_s(n_s)$ is the expected width of the
equal-tailed interval of $\operatorname{Beta}(a_s+y,b_s+n_s-y)$ under the
Beta--binomial predictive of $y$. Candidate totals lie in
\begin{equation}
\mathcal N_{t+1}
\subseteq
\left[
\max\left\{n_{\min},\sum_s n_{s,\min}\right\},
\min\left\{n_{\max},\sum_s n_{s,\max}\right\}
\right]\cap\mathbb N.
\label{eq:cs-admissible-totals}
\end{equation}
The selected total is allocated starting at the stratum floors, and each further
specimen goes to an eligible stratum with the largest reduction
$\overline W_s(n_s)-\overline W_s(n_s+1)$, with ties broken by a fixed order.
The Beta proxy sets the allocation only and does not enter the assurance
calculation.

\paragraph{Assurance.}
To reduce cost, the predictive ensemble is first reduced to $N_{\mathrm{sup}}$
equally weighted particles by systematic resampling. The accuracy of the
assurance estimate is therefore governed by $N_{\mathrm{sup}}$ and
$R_{\mathrm{MC}}$, not by the number of filter particles $N$. For a total $n$, each of
$R_{\mathrm{MC}}$ replicates draws one support particle $J_r$ and then pooled
four-cell counts
$\mathbf Y_{r}\sim\operatorname{Multinomial}(n;\bar{\mathbf q}_{t+1}^{(J_r)})$
from Eq.~\eqref{eq:cs-pooled-cells}. Each support particle is then reweighted
by the tempered likelihood of these hypothetical counts,
\begin{equation}
w^{(i)}(\mathbf Y_r,n)
\propto
\prod_{a,b}\left(\bar q_{t+1}^{ab,(i)}\right)^{\zeta Y_r^{ab}}.
\label{eq:cs-hypothetical-weights}
\end{equation}
The hypothetical width $W_r$ is the difference between the weighted 95\% and
5\% quantiles of $\psi_{t+1}^{(i)}$, and assurance is estimated by the fraction
of replicates with $W_r\le w^*$, with conditional variance
$\widehat A(1-\widehat A)/R_{\mathrm{MC}}$. The draws use the untempered
multinomial, while the hypothetical update uses the same tempering as the
filter. Because the update reweights a fixed support, weights can concentrate
at large $n$ and make the interval too narrow, which would make the estimated
assurance optimistic. The replay did not monitor this.

\subsection*{Retrospective replay protocol and performance measures}
\phantomsection
\label{app:replay-protocol}

Before the first decision round the filter assimilated the observed burn-in
data. At each decision time $t$ the replay propagated the filtering
distribution given $h_t^{\mathrm{cf}}$ to round $t+1$, built the allocation
family and selected $\mathbf n_{t+1}^*$ without using round-$t+1$ data, drew
the counterfactual specimens by simple random sampling without replacement
within each stratum's pool, truncated at the pool size, and assimilated their
paired results to form $h_{t+1}^{\mathrm{cf}}$. Each region was replayed with
three random seeds. Because every drawn observation changes all later
decisions, replay variability can be estimated only by repeating the whole
sequence.

Performance was measured against the full observed crude regional proportion
\begin{equation}
\widetilde p_{u,N}^{\mathrm{obs}}
=
\frac{\sum_{s=1}^S y_{s,u,N}^{\mathrm{obs}}}
{\sum_{s=1}^S n_{s,u,N}^{\mathrm{obs}}}.
\label{eq:app-observed-crude-proportion}
\end{equation}
RMSE was the root mean squared difference between the filtering mean and
$\widetilde p_{u,N}^{\mathrm{obs}}$ over rounds with data, mean width the
average width of the 90\% filtering interval, and empirical inclusion the
fraction of rounds in which that interval contained
$\widetilde p_{u,N}^{\mathrm{obs}}$. These measures describe agreement with the
historical series rather than error against the estimand, for two reasons. The
reference contains the counterfactual subsample, and it weights strata by donor
counts rather than by the population weights $\omega_s$ that define $\psi_u$.
Donors are not distributed across age and sex in population proportions, so
the difference
$\sum_s(\omega_s-n_{s,u,N}^{\mathrm{obs}}/\sum_rn_{r,u,N}^{\mathrm{obs}})\pi_{s,u}$
enters the RMSE as a bias, and empirical inclusion should not be compared with
$1-\tau$.

\end{document}